\documentclass[pre,aps,twocolumn,showpcs,floatfix,amsmath,amssymb]{revtex4}
\usepackage{amsmath}
\usepackage{graphicx,dcolumn,bm,subfigure} 
\usepackage{bm}        
\usepackage{mathrsfs} 
\usepackage{color}
\usepackage[utf8]{inputenc}
\usepackage{amssymb}

\def\3dots{\:\raisebox{-0.5ex}{$\stackrel{\textstyle.}{:}$}\:}

\def\beq{\begin{equation}}
\def\eeq{\end{equation}}
\def\bea{\begin{eqnarray}}
\def\eea{\end{eqnarray}}

\begin{document}
\title{Constraint percolation on hyperbolic lattices}
\author{Jorge H. Lopez$^1$ and J. M. Schwarz$^{2,3}$}
\affiliation{$^1$Facultad de Ciencias Exactas y Naturales, Universidad de Ibagu\'{e}, Ibagu\'{e}, Colombia}
\affiliation{$^2$Department of Physics, Syracuse University, Syracuse, NY 13244, USA}
\affiliation{$^3$Syracuse Biomaterials Institute, Syracuse, NY 13244}  
\begin{abstract}
Hyperbolic lattices interpolate between finite-dimensional lattices and Bethe lattices and are interesting in their own right with ordinary 
percolation exhibiting not one, but two, phase transitions. We study four constraint percolation models---$k$-core percolation (for 
$k=1,2,3$) and force-balance percolation---on several tessellations of the hyperbolic plane. By comparing these four different models, 
our numerical data suggests that all of the $k$-core models, even for $k=3$, exhibit behavior similar to ordinary percolation, while the 
force-balance percolation transition is discontinuous. We also provide a proof, for some hyperbolic lattices, of the existence of a critical 
probability that is less than unity for the force-balance model, so that we can place our interpretation of the numerical data for this 
model on a more rigorous footing. Finally, we discuss improved numerical methods for determining the two critical probabilities on the 
hyperbolic lattice for the $k$-core percolation models.
\end{abstract}

\maketitle

\section{Introduction}
Geometry plays a key role in driving physical processes in such different physics fields as relativity, cosmology, quantum field theories, 
and condensed matter \cite{Carroll, Fleury, Maia,Badyopadhyay, Nelson, Kamien, Kallosh}. In condensed matter systems, the effect of 
geometry on the nature of a phase transition is of particular interest \cite{Kung, Ruppeiner}. For example, hyperbolic spaces possessing 
a constant negative curvature of $-1$ have been recently applied to several condensed matter models, namely the Ising model 
\cite {Rietman, Chris, Gandolfo, Krcmar1, Krcmar2, Shima, Ziff_Hyperbolic} and percolation \cite{Baek2, sausset2, Baek1, benjamini}. 

Why consider hyperbolic spaces? Hyperbolic geometry connects to properties of mean field theory as studied on Bethe lattices with the 
same nonvanishing ratio of surface to volume of compact structures as the size of the lattice scales to infinity \cite{Shima, Krcmar2}. 
And yet there are loops at all length scales as is the case with Euclidean lattices. Accordingly, hyperbolic lattices provide a test bed for 
studying phase transitions in a geometry that interpolates between Bethe lattices and Euclidean lattices. Hyperbolic lattices are also 
interesting from a glassy physics perspective because they provide a natural mechanism in two dimensions to frustrate global crystalline 
order and allow for a more tractable model to study the glass transition and jamming in two dimensions~\cite{Kivelson, Modes} .

A hyperbolic lattice is a tessellation of the hyperbolic plane, usually denoted by the so called Schl\"{a}fli symbol  $\{P,Q\}$, where regular 
polygons of $P$ sides tile the plane so that $Q$ of these polygons meet at each vertex ~\cite{Coxeter}, and $P$, $Q$ satisfy the relation 
\begin{equation}
(P-2)(Q-2)>4
\end{equation}
It should be noted that (1) Euclidean lattices satisfy the equation $(P-2)(Q-2)=4$ and (2) for lattices on the elliptic plane, the relation 
$(P-2)(Q-2)<4$ holds~\cite{Raffield}. Therefore, the elliptic and Euclidean planes admit just a finite number of tessellations, while the 
hyperbolic plane is much more richer admiting an infinite number. We will use the Poincare disk respresentation of the hyperbolic plane, 
which is the unit radius disk with its respective metric~\cite{Anderson}. 

\begin{figure}[hb!]
  \centering
    \includegraphics[width=0.35\textwidth]{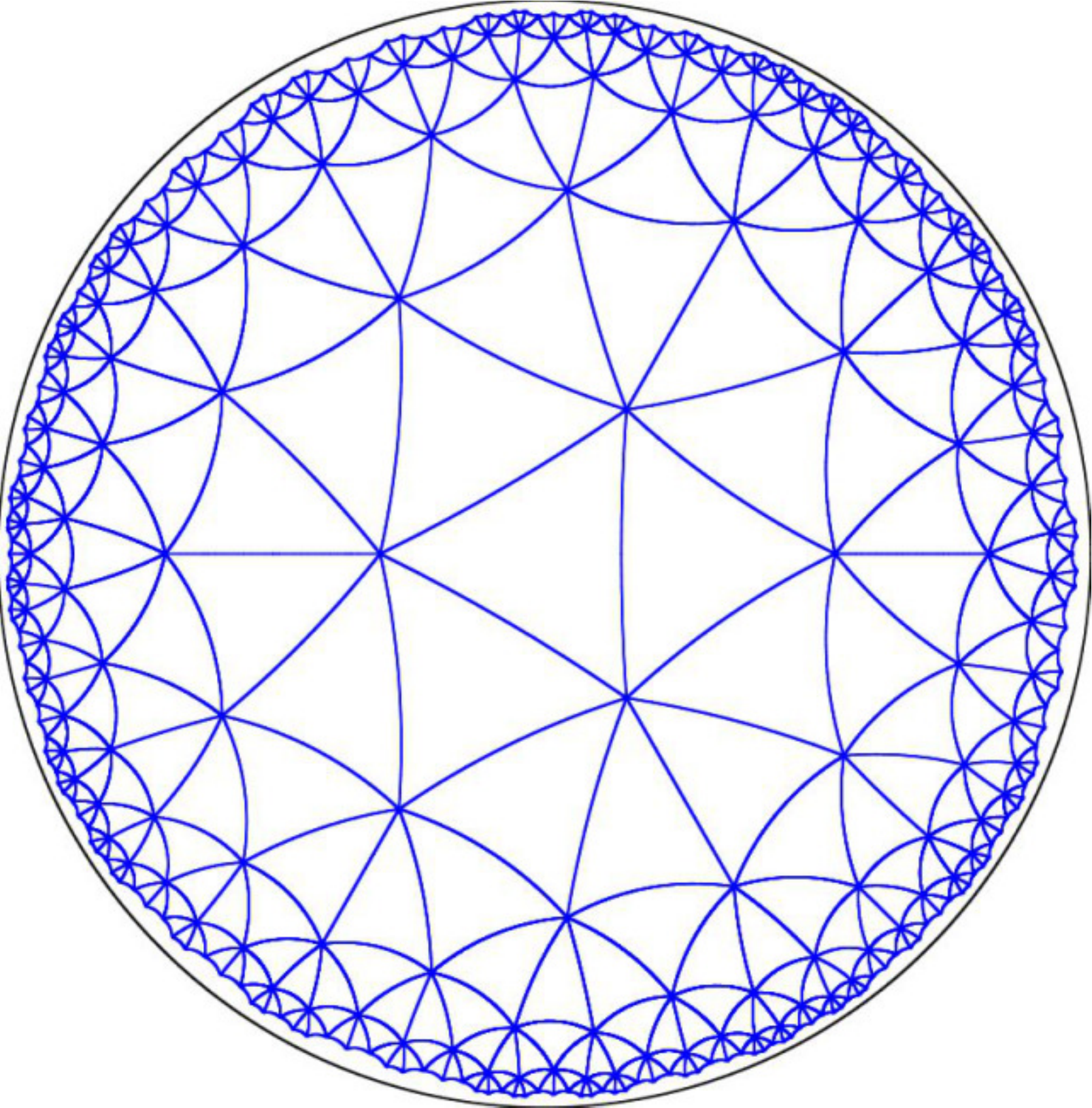}
	\caption{$\{3,7\}$ tessellation on the Poincare disk.}
	\label{Tes3_7}
\end{figure}

We will work with several hyperbolic tessellations, an example of which is seen in Figure~\ref{Tes3_7}, to study $k$-core~\cite{chalupa} 
and force-balance~\cite{schwarz2} percolation models and explore the nature of their transition. $k$-core percolation is constraint 
percolation model where occupied sites having less than $k$ occupied neighboring sites are pruned starting with an initial random and 
independent occupation of sites. This pruning is done consecutively until all occupied sites have at least $k$ occupied neighboring sites. 
This constraint imposes the scalar aspect of the local Hilbert stability criterion for purely repulsive particles in $d+1$ dimensions~\cite{Alexander}. 
For example, in two-dimensions, a particle must be surrounded at least three particles (the scalar aspect). In addition, 
at least three of these neighboring particles must enclose a particle within a triangle so that forces balance and each particle is locally 
mechanically stable. In mean field, $k$-core percolation resembles some properties of a mixed phase transition~\cite{schwarz2}, 
i.e. discontinuity in the order parameter and a diverging length scale, as in the jamming transition~\cite{epitome}. And yet, $k$-core 
percolation on Euclidean lattices appears exhibit either a continuous phase transition in the same universality class as ordinary percolation~\cite{brazilians}, 
or no transition~\cite{vanEnter}. So we ask the question: What is the nature of the $k$-core percolation transition on 
hyperbolic lattices? Will the transition behave more like what is computed on the Bethe lattice, or not?

As you will soon discover, many of the numerical techniques developed
for the analysis of the phase transition in ordinary percolation are
not as readily applicable on hyperbolic lattices given the strong
boundary effects, which makes the above questions slightly difficult
to answer. There is also the possible complication that there are two
phase transitions, as has been demonstrated for ordinary percolation---one
transition at the onset of many spanning clusters touching the boundary and a
second transition at the onset of all of the spanning merging into just one spanning
cluster~\cite{benjamini}.  Ref.~\cite{sausset} asked the above questions
for the $k=3$ case and concluded, based on a conjecture and on
numerical evidence, that the mixed nature of the $k=3$-core
percolation transition on the Bethe lattice was robust on the
hyperbolic lattice. In light of more recent work identifying crossing
probabilities on the hyperbolic lattice for ordinary
percolation~\cite{Ziff_Hyperbolic}, we revisit the above questions for $k=3$-core
percolation and analyze the other $k$-core models as well. 

Given the numerical intricacies, as a means 
of comparison, we also investigate force-balance percolation on hyperbolic lattices. Force-balance percolation is another constraint percolation 
model~\cite{schwarz2} in which each site is first occupied with some occupation probability $p$. Then, if an occupied site is not surrounded by 
neighboring occupied sites such that the local Hilbert stability
criterion cannot be guaranteed, the occupied site is removed from the
lattice.  Again, the local Hilbert stability criterion ensures for a
purely repulsive system that forces can be balanced and that each
particle be locally mechanically stable. This 
procedure is repeated until all occupied sites obey the occupation constraint. This model was studied in two- and three-dimensions~\cite{schwarz1, schwarz2}. Numerical simulations suggested strongly signs of a discontinuous transition in the standard order parameter (i.e. the 
fraction of sites participating in the spanning cluster), which also occurs in jamming.  Numerical simulations also suggested that 
there exists a correlation length scale diverging faster than any
power-law, which is different from jamming where numerics suggests a  
more standard power-law diverging correlation length~\cite{ohern1}. 

So we expect force-balance percolation to exhibit a mixed transition on various Euclidean lattices. A mean field theory of force-balance 
percolation is not possible because spatial information is encoded in the constraints.  Studying force-balance percolation on hyperbolic 
lattices will allow us to work towards a mean field theory without giving up the spatial constraints.  We expect force-balance percolation 
to exhibit a discontinuous percolation transition on hyperbolic
lattices since it already appears to be in the presence of many
loops~\cite{schwarz1,schwarz2}.  Perhaps, however, the diverging length
scale on the hyperbolic lattice will be a power-law, as opposed to
faster than a power-law on the Euclidean lattice. In any event, the
discontinuity in the onset of the spanning cluster should give us something to compare against when trying to determine whether or not $k$-core percolation exhibits a discontinuous 
transition on hyperbolic lattices. 

The remainder of this manuscript is organized as follows: We will study several properties of $k$-core percolation models for $k=1,2,3$, and 
force-balance percolation on hyperbolic tessellations. We present in Section 2 details of the hyperbolic lattice and various percolation 
algorithms. In Section 3, we present a theoretical proof that the threshold for force-balance percolation is strictly less than one for most of 
the tessellations. This section is a bit technical and can be skipped by the reader should their interest be more in the nature of the phase 
transition. We present our numerical results in Section 4, where we
study the crossing probability and other measurements. We summarize and discuss the implications of our results in Section 5.

\section{Model and methods}
The key step in the simulation process is to construct a hyperbolic
lattice. We do this by implementing the algorithm described in detail
in Ref.~\cite{Dunham}. In the construction of a $\{P,Q\}$ hyperbolic
lattice, where again, $P$ denotes the number of sides of each polygon
and $Q$ denotes how many polygons meet at a vertex, the central
polygon is built first, and this is the first layer. Then, by
translations and rotations of the central polygon, the second layer is
built. This process is followed recursively until a desired number of
layers is constructed. An $l$-layer is composed of those polygons that
do not belong to an $m$-layer for $m<l$ and share an edge or vertex
with a polygon in the $(l-1)$-layer. The algorithm makes use of the
\textit{Wierstrass model} for hyperbolic geometry, where points lie on
the upper sheet of the hyperboloid,
$x^{2}+y^{2}-z^{2}=-1$. Consequently, rotations and translations are
given by $3\times3$ Lorentz matrices. The Wierstrass model is related to the Poincare model through the stereographic projection toward the point $(0,0,-1)^{t}$ given by
\begin{equation}
\begin{pmatrix}
x \\
y \\ 
z 
\end{pmatrix}
\longrightarrow \dfrac{1}{1+z}
\begin{pmatrix}
x \\ 
y \\ 
0 
\end{pmatrix}.
\end{equation}

The exponential growth of number of vertices with respect to the number of layers constrains severely the number of layers used in the simulations. Typically we simulate around 10 layers. This is comparable to the recent work by Gu and Ziff studying ordinary percolation on hyperbolic lattices~\cite{Ziff_Hyperbolic}.  Recent work on implementing periodic boundary conditions in certain tilings may ultimately be investigated~\cite{sausset}. However, the sets of hyperbolic tillings that can be used using the methods in Ref.~\cite{sausset} have less than 30000 sites due to a lack of knowledge of all possible normal subgroups of a given Fuchsian group.

Once a tessellation is created, each of its sites are occupied with
probability $p$. For $k$-core percolation, we then recursively remove
any occupied site (excluding boundary sites) that has less than $k$
occupied neighboring sites. For force-balance percolation, we
recursively remove any occupied sites (excluding boundary sites) that
are not enclosed by a triangle of neighboring occupied sites,
i.e. those sites that are not locally mechanically stable. We do this until all occupied sites obey 
the imposed constraint. We have numerically tested on around one million runs, that the order 
in which we check the force-balance constraint does not affect the final configuration, i.e. that 
the model is abelian. It has been also argued that the $k$-core model is abelian~\cite{Abelian_kcore}. 

\begin{figure}[h]
  \centering
    \includegraphics[width=0.35\textwidth]{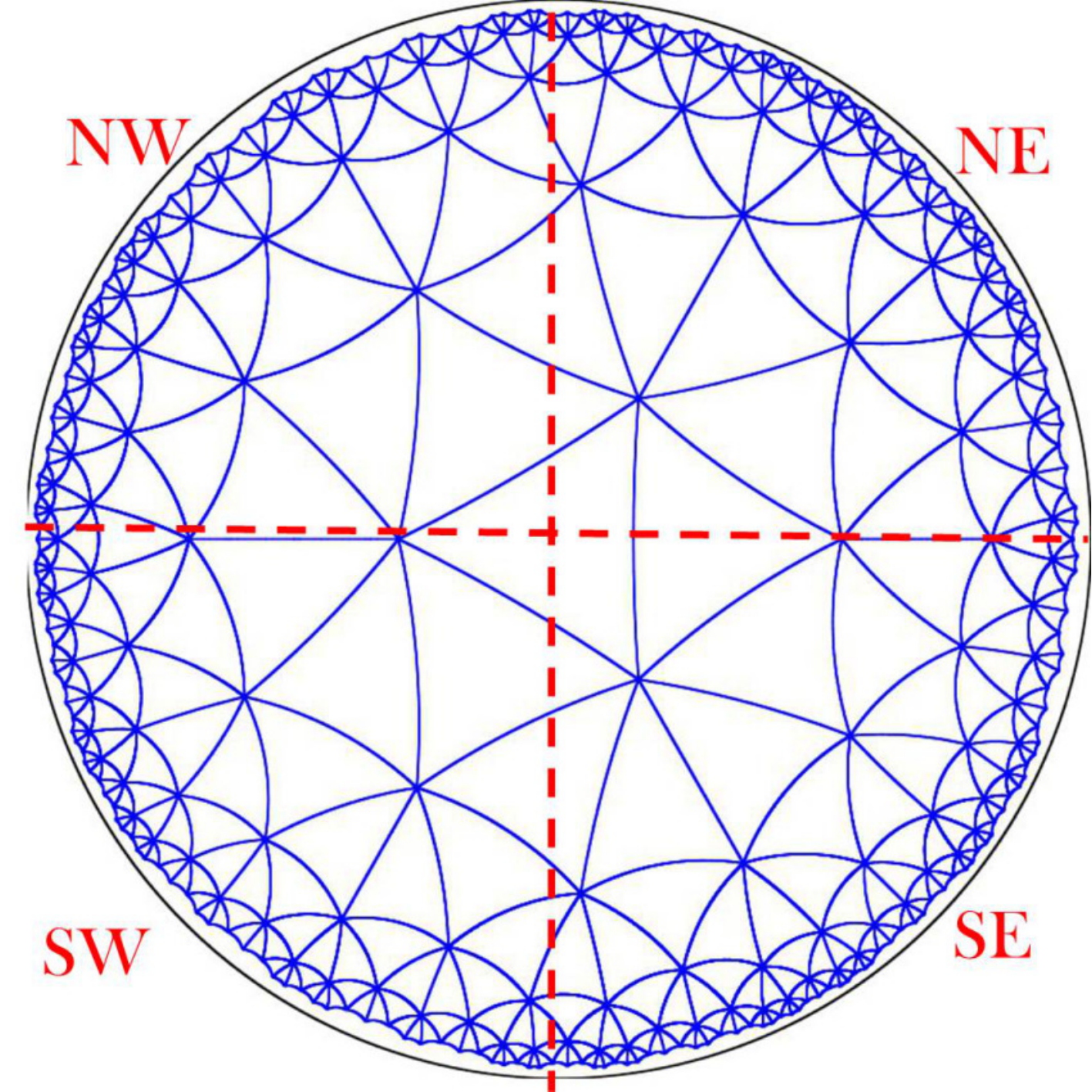}
	\caption{$\{3,7\}$ tessellation on the Poincare disk with the four boundary regions.}
	\label{Tes3_7_Quarters}
\end{figure}

We then use the Hoshen-Kopelman algorithm to identify the clusters and
their respective sizes. To determine if a cluster is spanning, we
break up the lattice into four cardinal regions: $NE$, $NW$, $SW$, and
$SE$. See Figure ~\ref{Tes3_7_Quarters}. We regard the cluster as
percolating, or spanning, when it connects either $NE$ and $SW$ sites
or $NW$ and $SE$ sites, as in Ref.~\cite{Ziff_Hyperbolic}. We then
measure the probability to span, or cross for an occupation
probability $p$ and denote it $R(p)$. We also measure a quantity
defined as $S_{1}/N$, where $S_{1}$ is the size of the largest cluster
and $N$ the total number of sites. This quantity resembles the order
parameter and, therefore, determines the continuity/discontinuity of
the onset of the transition(s), i.e. should it increase from zero
continuously as the occupation probability $p$ is increased, then the
transition is continuous. We also measure the number of times we check
the lattice to cull occupied sites not obeying their respective
constraints, i.e. the culling time. This \textit{culling time} tends
to diverge near transition on Euclidean lattices~\cite{schwarz2}.

\section{Proof of $p_{FB}<1$ for some hyperbolic tillings}
\label{sec:TheoreticalProof}
It has been established that there exists two critical percolation probabilities, $p_{l}$ and $p_{u}$ for ordinary 
percolation on hyperbolic tillings ~\cite{Czajkowski, Baek1,Lee}. For the force-balance model, however, it seems 
there is just one critical percolation probability, according to the
results presented later, demonstrating the emergence of a percolating cluster. Let us call this 
probability, $p_{FB}$, the probability above which there is always a percolating cluster. It is possible to prove 
that $p_{FB}<1$ for some hyperbolic tillings $\{P,Q\}$. The proof follows two steps: 
\begin{enumerate}
	\item First establish the existence of trees on a tessellation $\{P,Q\}$ with a certain connectivity that 
	depends on the parity of $Q$. For $Q$ even we demand a connectivity $z=6$ and for $Q$ odd, $z=5$. 
	\item We apply a well known result of $k$-core percolation on trees, i.e. that the critical percolation is 
	less than one when $k<z$~\cite{chalupa}. For our purposes, we require $k=5$ for tessellations of $Q$ even and 
	$k=4$ for $Q$ odd. Accordingly, we show that sites on a
        percolating cluster for the $k=5$-core model on 
	the $z=6$ trees, and $Q$ even, satisfy the occupation
        constraints of the force-balance percolation model. Similarly,
        for the $Q$ odd case, we study the $k=4$-core model on $z=5$
        trees. 
\end{enumerate}

\begin{figure}[hb!]
  	\centering
    \includegraphics[width=0.35\textwidth]{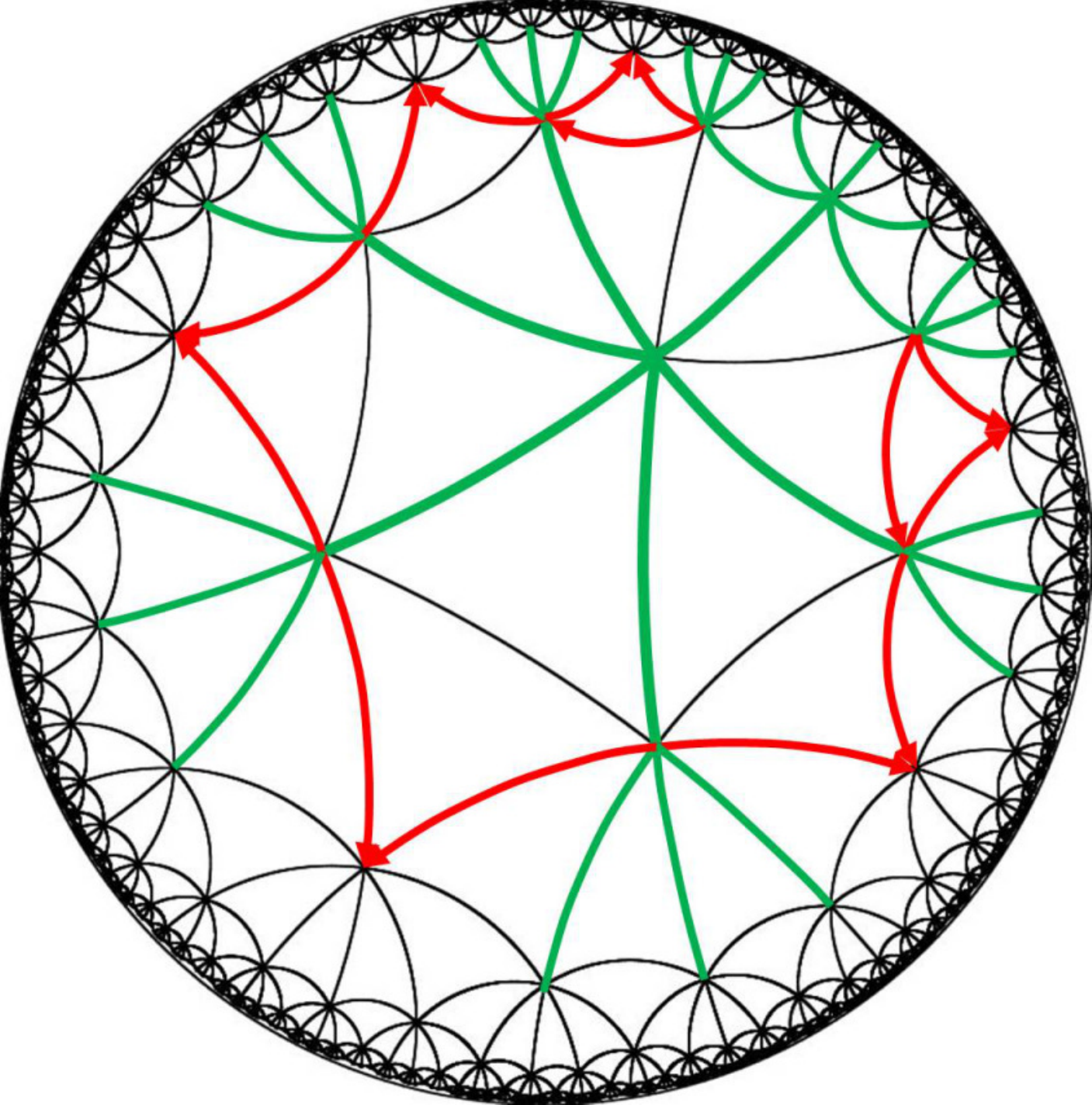}
  	\caption{One cannot embed a tree of connectivity $z=6$  on the
          $\{3,8\}$ tessellation due to the lack connections.}
	\label{Tes_3_8}
\end{figure}

\begin{figure}[hb!]
  \centering
    \includegraphics[width=0.35\textwidth]{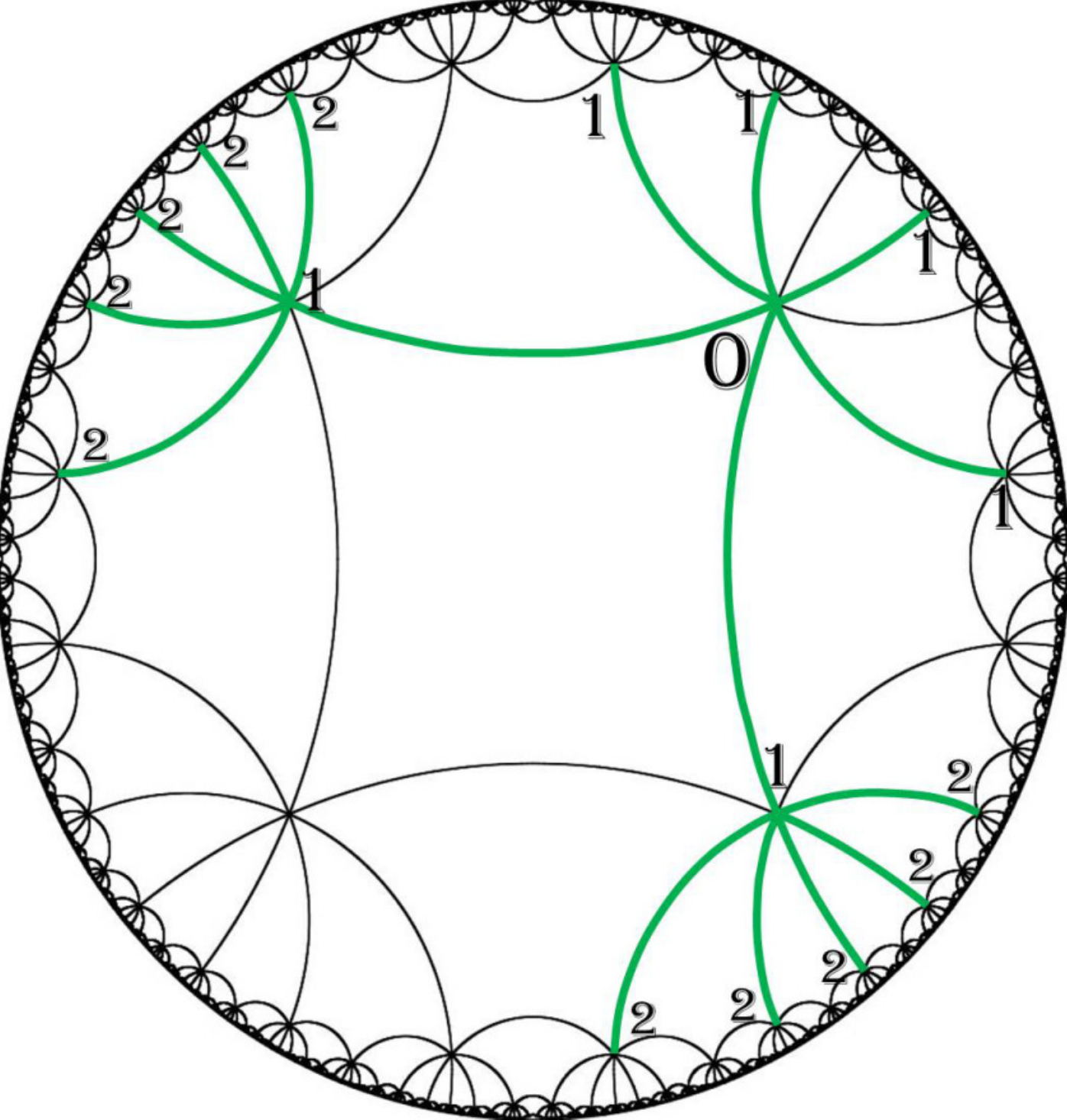}
  \caption{Tessellation $\{4,8\}$ enables the construction of trees of connectivity $z=6$}
	\label{Tes4_8}
\end{figure}

Let us prove each of these items in due order. First, we need to show the existence of trees of connectivity $z=6$ and $z=5$ for $Q$ even and odd, 
respectively. Let us suppose $Q$ is even. It is easy to see that $z=6$
trees cannot be built when $Q=4, 6$ 
as there is not enough ``space" to build trees given the eventual
overlaps. The case $Q=8$ is more interesting. The tessellation
$\{3,8\}$ does not admit a tree construction due to overlaps, as
illustrated in Fig.~\ref{Tes_3_8} where red arrows shows some of those
positions at which the initial tree (green) eventually contains
overlaps. However, $z=6$ trees can be built on the tessellation $\{4,8\}$. To see this, we choose a site which we call the 
$0th$-generation. The first generation are the neighbors of such a site. The $nth$-generation will be formed by those site neighbors of the $(n-1)th$-generation that do not belong to a $kth$-generation where $k<n$. 
This is illustrated in Fig.~\ref{Tes4_8}. By construction, between two adjacent $1st$-generation sites on 
the $z=6$ tree there is one $2nd$-generation site which does not belong to the tree. Now between the 
closest offspring of those $1st$-generation sites which are $2nd$-generation sites belonging to the three, there 
are six $3rd$-generation sites not belonging to the tree. By construction, such trees can be expanded without 
overlapping so that they, indeed, remain trees.

For $P>4$ we have more vertices in each layer, which gives more space to build trees, and the same 
construction holds. Accordingly, we can build $z=6$ trees on the
tessellation $\{P,8\}$ when $P>3$. Likewise, it can be checked that for any $P$, $Q$ even and $Q>9$, it is possible to build a tree of connectivity $z=6$.  
Analogously, trees of connectivity $z=5$ can be built on tessellations $\{P,7\}$
where $P>3$, and for any tessellation $\{P,Q\}$ where $Q>8$ is odd. 

In summary, those trees necessary for our
proof can be built on any tessellation $\{P,Q\}$ as long as $Q>8$ and for the tessellations $\{P,7\}$, $\{P,8\}$
as long as $P>3$.

\begin{figure}[ht!]
	\centering
	\subfigure{
		\includegraphics[width=0.35\textwidth]{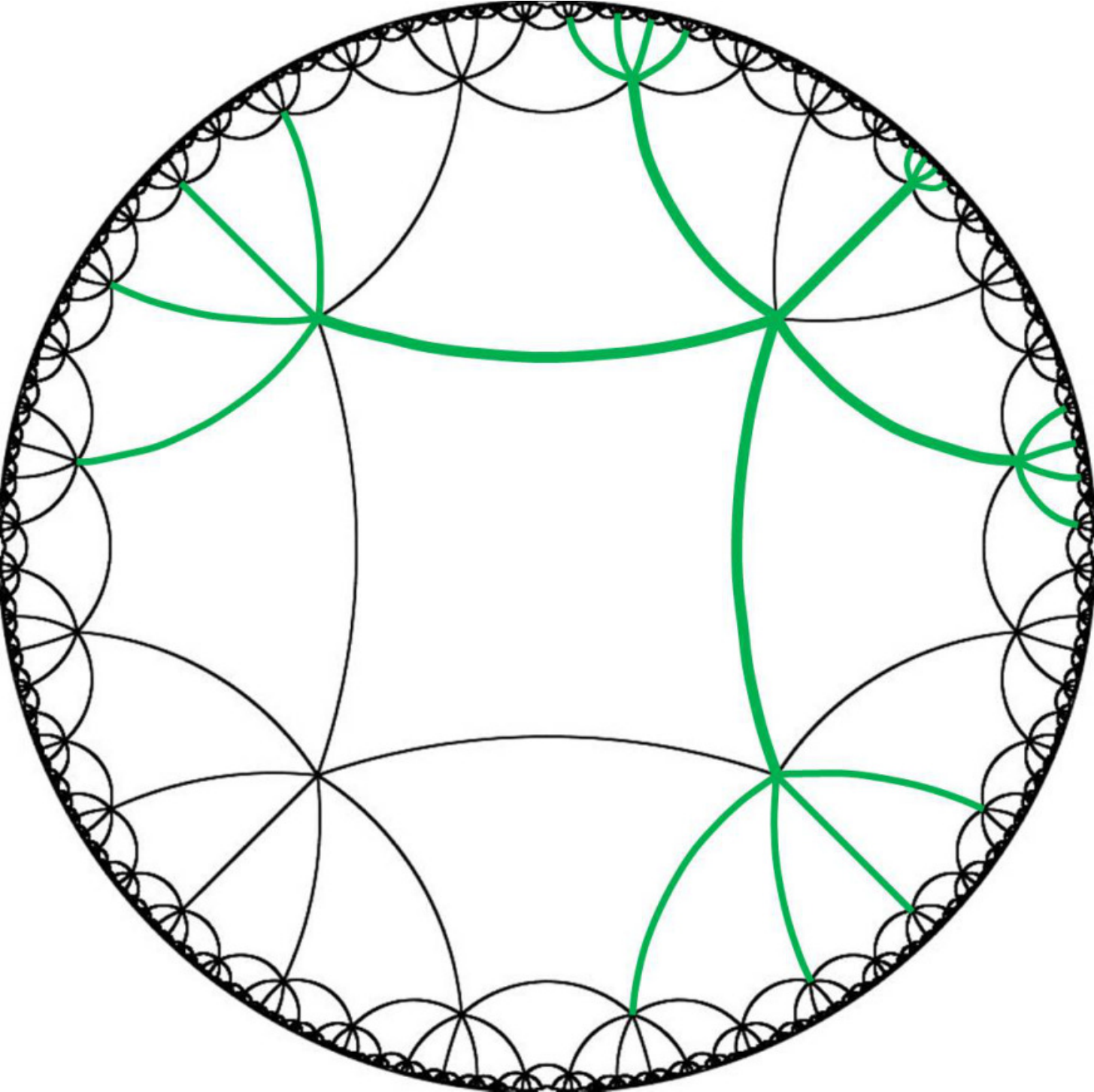} 
		\label{fig1:4_7}}		
	\subfigure{
		\includegraphics[width=0.35\textwidth]{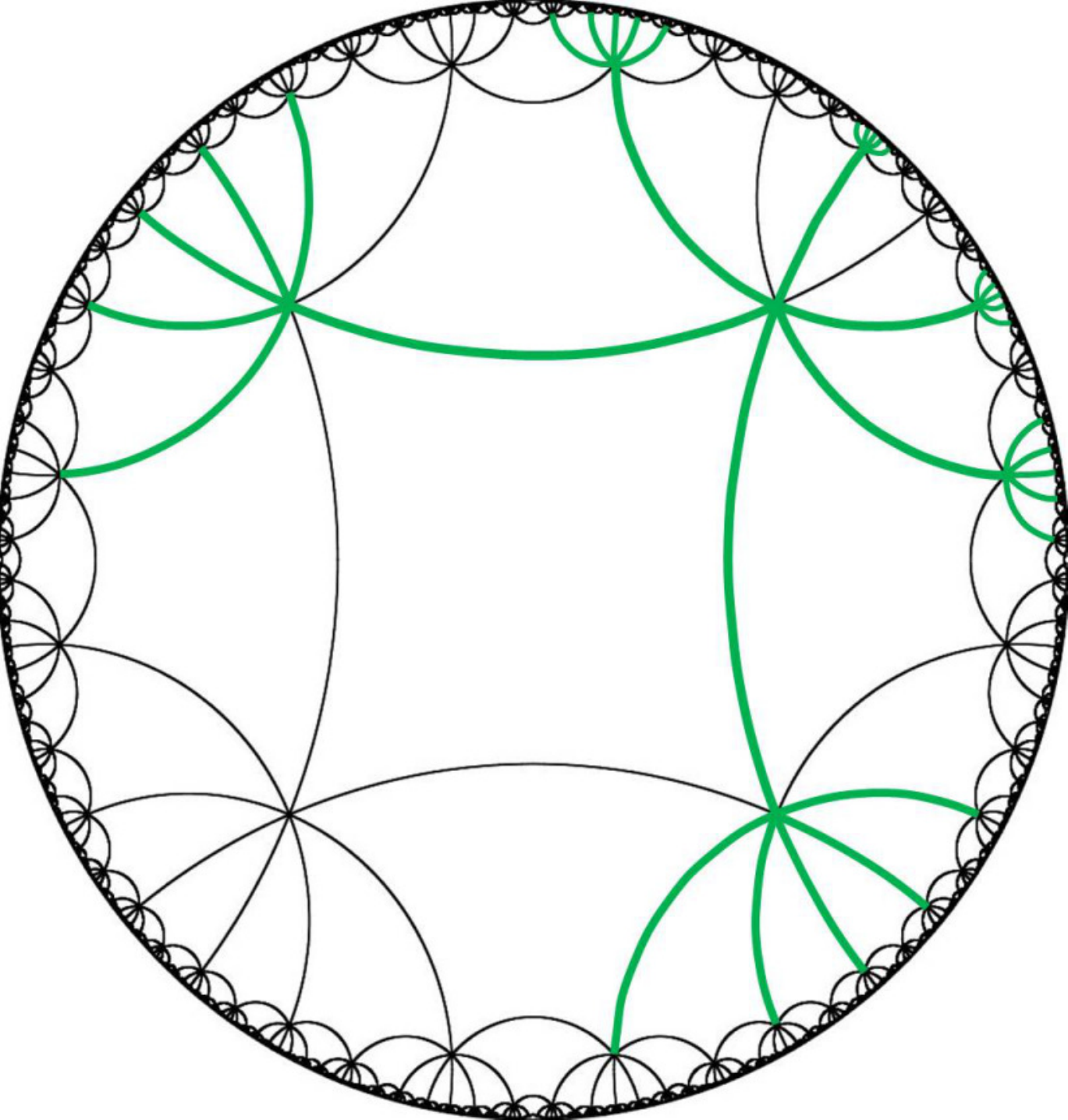} 
		\label{fig1:4_8}}	
	\caption{Tree construction on tessellations $\{4,7\}$ and
          $\{4,8\}$: Top: Tree of connectivity $z=5$ on the $\{4,7\}$
          tessellation. Bottom: Tree of connectivity $z=6$ on the $\{4,8\}$ tessellation.}
	\label{fig1:Tessellations}
\end{figure}

\begin{figure}
	\centering
	\hspace*{\fill}
		\includegraphics[width=0.2\textwidth]{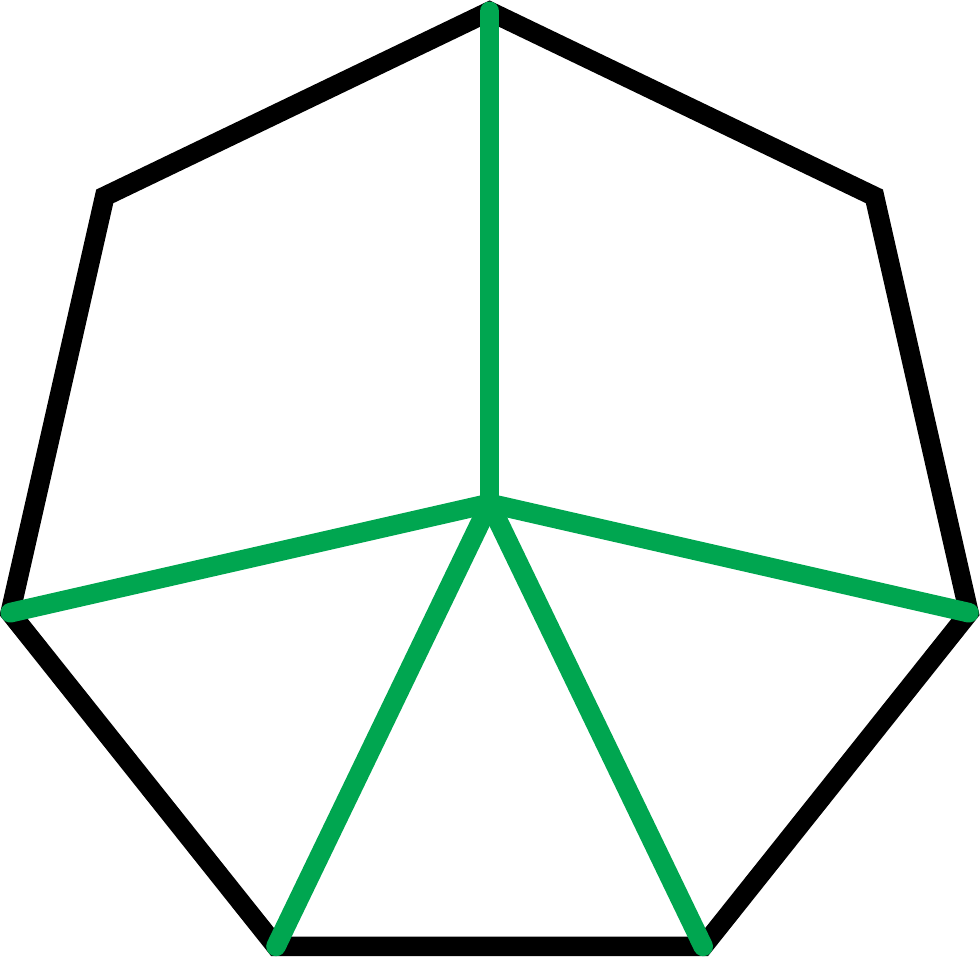} \hfill
		\includegraphics[width=0.2\textwidth]{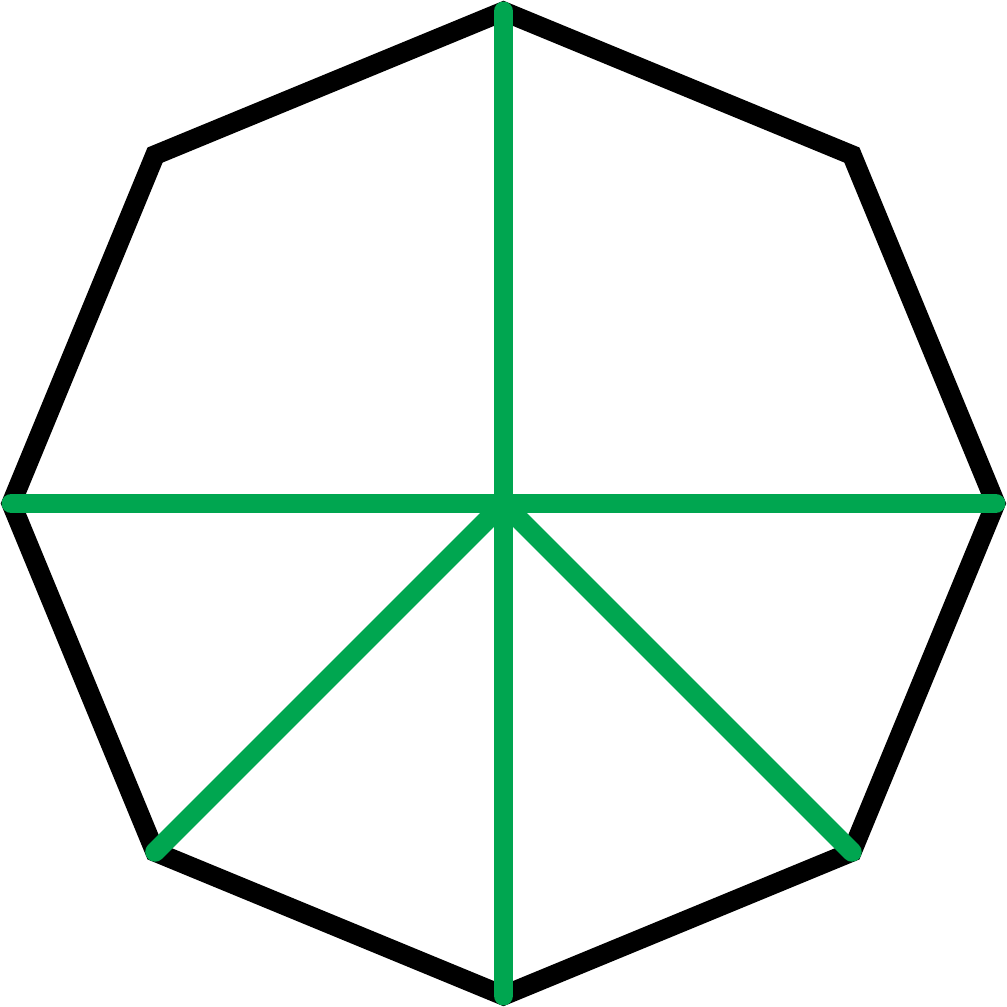}	
		\hspace*{\fill}
	\caption{Euclidean illustration of the central part of the
          trees on hyperbolic tessellations: Left: Euclidean
          illustration of the "central" part of the $z=5$ tree on
          tessellation $\{4,7\}$. Right: Euclidean illustration of the "central" part of the $z=6$ tree on tessellation $\{4,8\}$.}
	\label{fig1:Polygons}
\end{figure}

\begin{figure}[hb!]
	\centering
	\subfigure{	
		\includegraphics[width=0.3\textwidth]{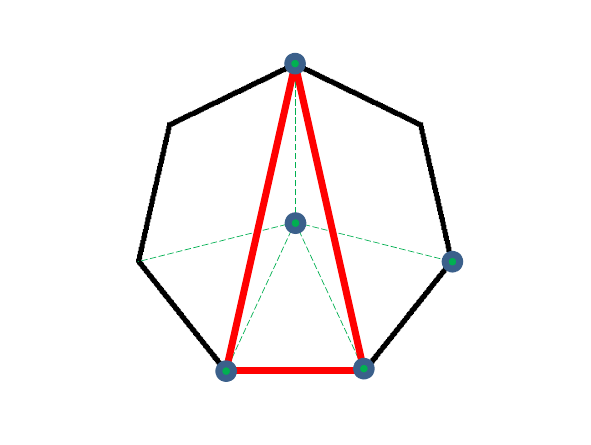}
		\label{fig2:1}	}
	\subfigure{	
		\includegraphics[width=0.3\textwidth]{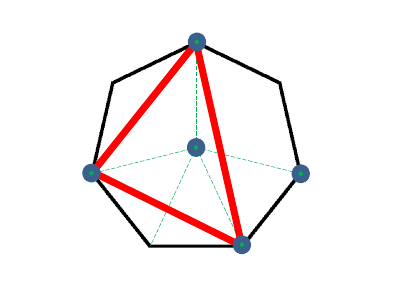}
		\label{fig2:2}	}
	\subfigure{	
		\includegraphics[width=0.3\textwidth]{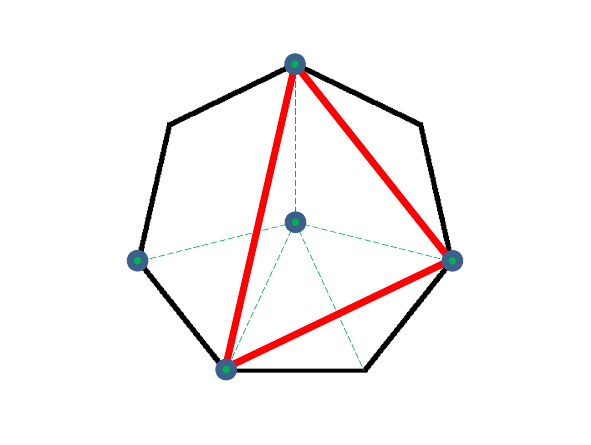}
		\label{fig2:3}	}
	\subfigure{	
		\includegraphics[width=0.3\textwidth]{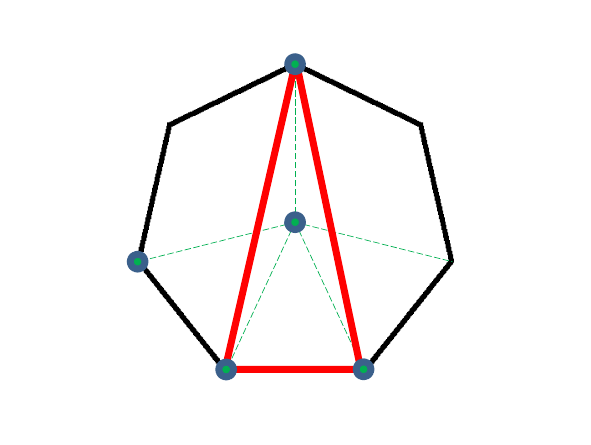}
		\label{fig2:4}	}
	\caption{Illustration of all the possible cases of occupation for a $k=4$-core cluster on a tree of connectivity $z=5$}	
	\label{Heptagon_FB}
\end{figure}

As for the second step in the proof, consider any site on the tree
built in step 1. A site of a $\{P,Q\}$ tessellation will be contained in a $Q$-gon as 
illustrated in Fig.~\ref{fig1:Tessellations}. Now let us take such
$Q$-gons in a Euclidean setting as illustrated in
Fig.~\ref{fig1:Polygons}. One of the neighbors of central site is
isolated from others. Let us call it the north 
neighbour, $NN$. It happens that any tree of connectivity $z=4$ ($Q$ even case) containing site $NN$ and imbedded in those 
trees of connectivity $z=5$, satisfy the force-balance constraint as
indicated in Fig.~\ref{Heptagon_FB}. In two-dimensions this constraint
is that every occupied site (particle) have at least three neighoring
occupied sites and at least three of these neighboring sites enclose
the occupied site in a triangle. This \emph{triangle} condition on the central site is preserved in 
the hyperbolic geometry given the function that relates those polygons in different geometries 
preserves topology. A similar proof applies to trees of 
connectivity $z=5$ embedded in trees of connectivity $z=6$ ($Q$ even case).

The above analysis holds for any site so we can always construct such a $Q$-gon with the same 
characteristics for any occupied site. Now let us call $p_{4c}$ the critical percolation 
probability for $k=4$-core percolation on trees of connectivity $z=5$ and $p_{5c}$ such probability for 
$k=5$-core percolation on trees of connectivity $z=6$. It follows from the discussion 
above that $p_{FB}<p_{4c}$ for $Q$ even, and $p_{FB}<p_{5c}$ when $Q$ is odd (search $p_{4c}$ and $p_{5c}$)
at least for those tessellations where we can make the tree
construction illustrated in Fig.~\ref{fig1:Tessellations}.  Since both
$p_{4c}$ and $p_{5c}$ are less than unity for the trees enumerated, $p_{FB}<1$.

\section{Results}
We work with tessellations $\{3,7\}$, $\{7,3\}$, and $\{4,7\}$, where
the first two tessellations are the most common
studied~\cite{Ziff_Hyperbolic,Baek1}.  We study $k=1,2,3$-core
percolation and force-balance percolation on such tessellations by
computing the crossing probability, $R$, the probability of participating
in the largest cluster, $P_{LC}$, and the culling time. 

\subsection{Crossing probability}
Ordinary percolation exhibits three phases on
the hyperbolic lattice~\cite{benjamini}. Specifically, for $p<p_{l}$
there is no percolating cluster, for $p_{l}<p<p_{u}$ there are
infinitely many percolating clusters, and for $p>p_{u}$ the infinitely
many percolating clusters merge to form just one percolating cluster. The existence of three phases is reflected in the crossing probability, $R(p)$. According to Ref.~\cite{Ziff_Hyperbolic}, as the number of layers tends to infinity, $R(p)$ tends to a function that in the intermediate phase is a straight line with finite slope in the infinite layer limit. If there is just one phase boundary, as with ordinary percolation on Euclidean lattices, then in the infinite system limit $R(p)$ jumps discontinously at the boundary from zero to one through some value of $R(p_c)$, the Cardy crossing value~\cite{cardy}, at the transition.  So there would be no finite slope region in the infinite system limit. 

\begin{figure}[h]
	\centering
	\subfigure{	
		\includegraphics[width=0.3\textwidth]{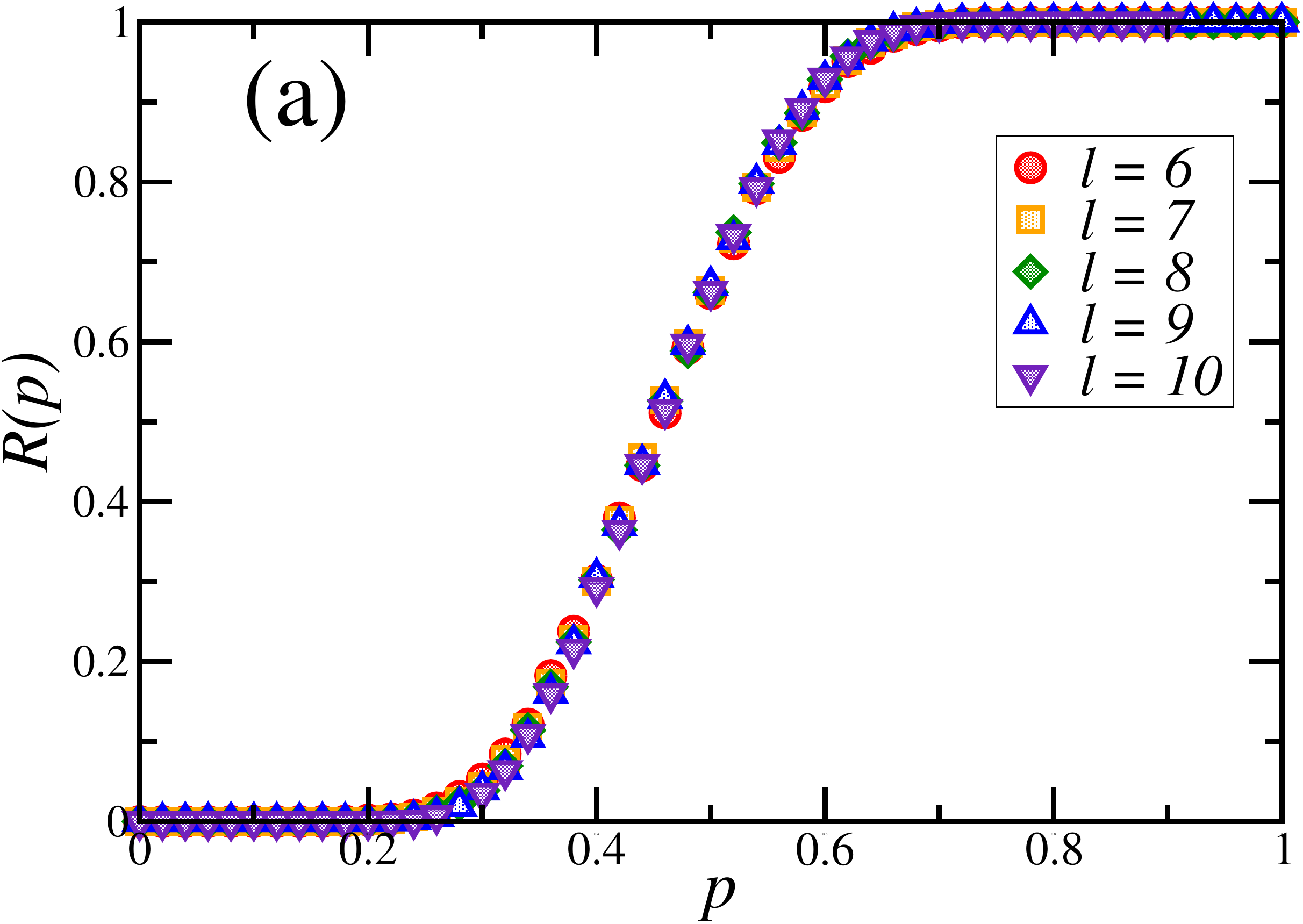}
		\label{CrossProb1:1}	}

	\subfigure{	
		\includegraphics[width=0.3\textwidth]{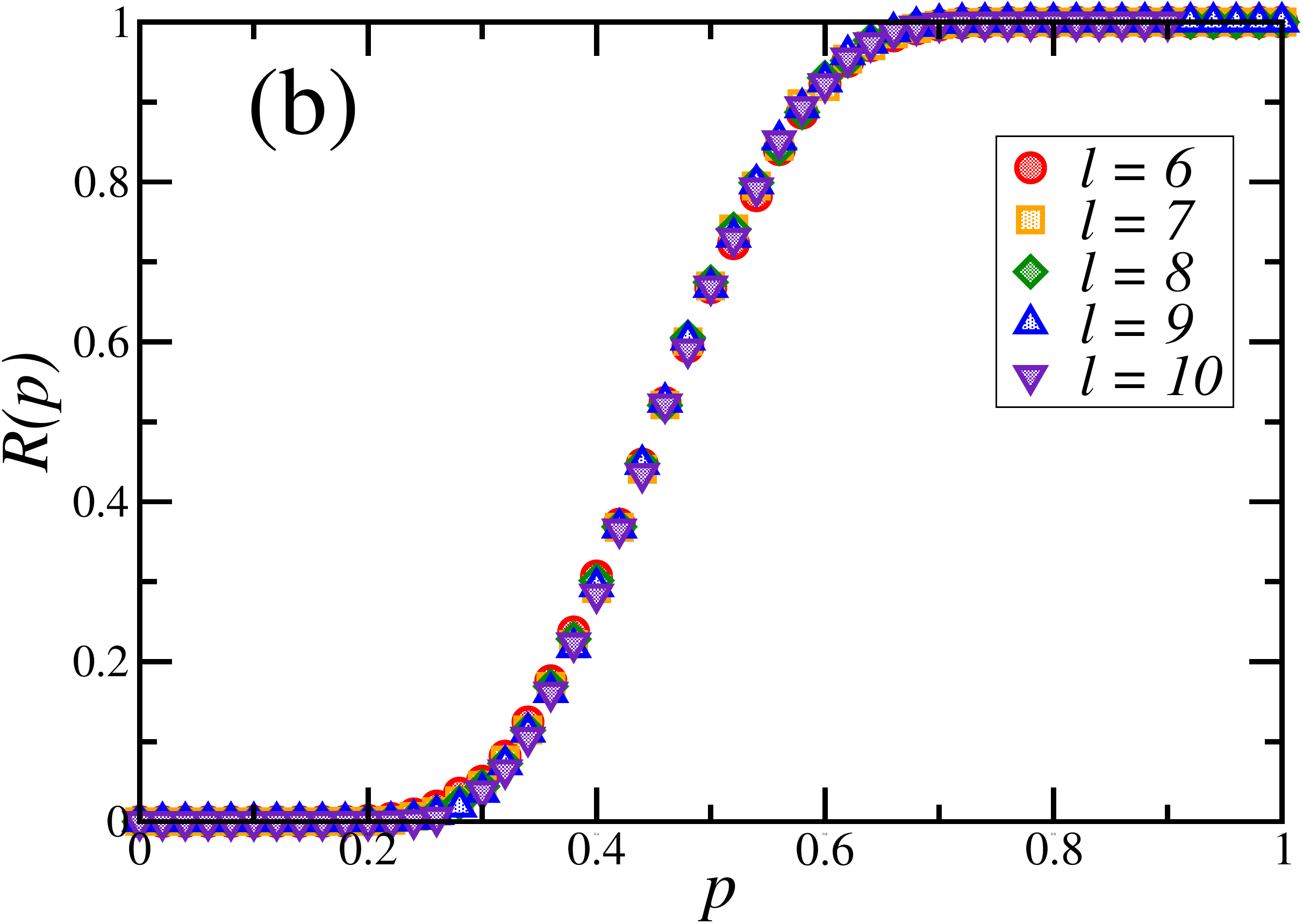}
		\label{CrossProb1:2}	}

	\subfigure{	
		\includegraphics[width=0.3\textwidth]{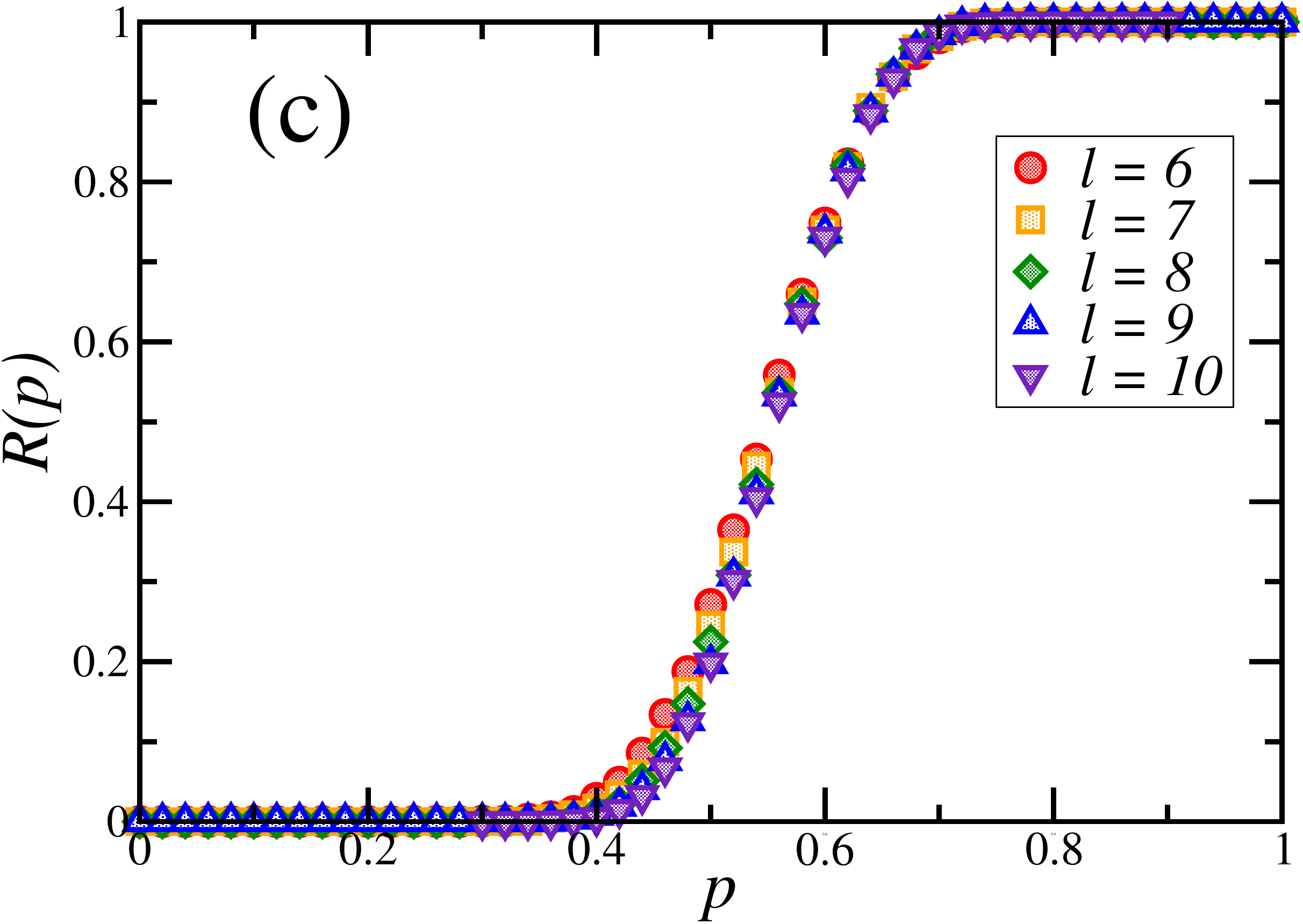}
		\label{CrossProb1:3}	}

	\subfigure{	
		\includegraphics[width=0.3\textwidth]{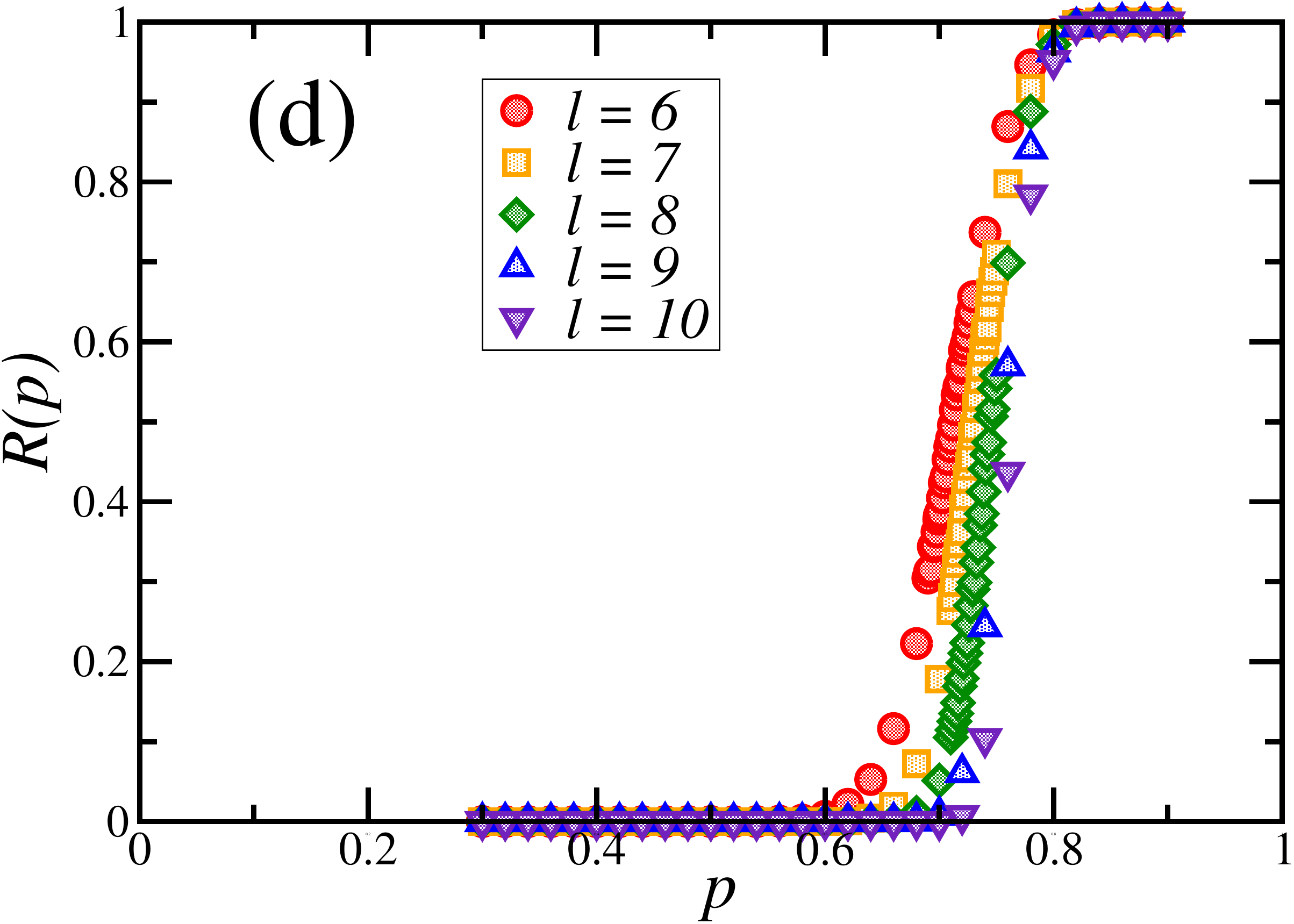}
		\label{CrossProb1:4}	}
	\caption{Crossing probability on $\{3,7\}$ tessellation for
          the different percolation models: (a) $k=1$-core, (b) $k=2$-core (c)
          $k=3$-core, (d) force balance.}	
	\label{CrossProb1}
\end{figure}       

\begin{figure}[hb!]
	\begin{center}
\vspace{0.5cm}
		\includegraphics[width=0.4\textwidth]{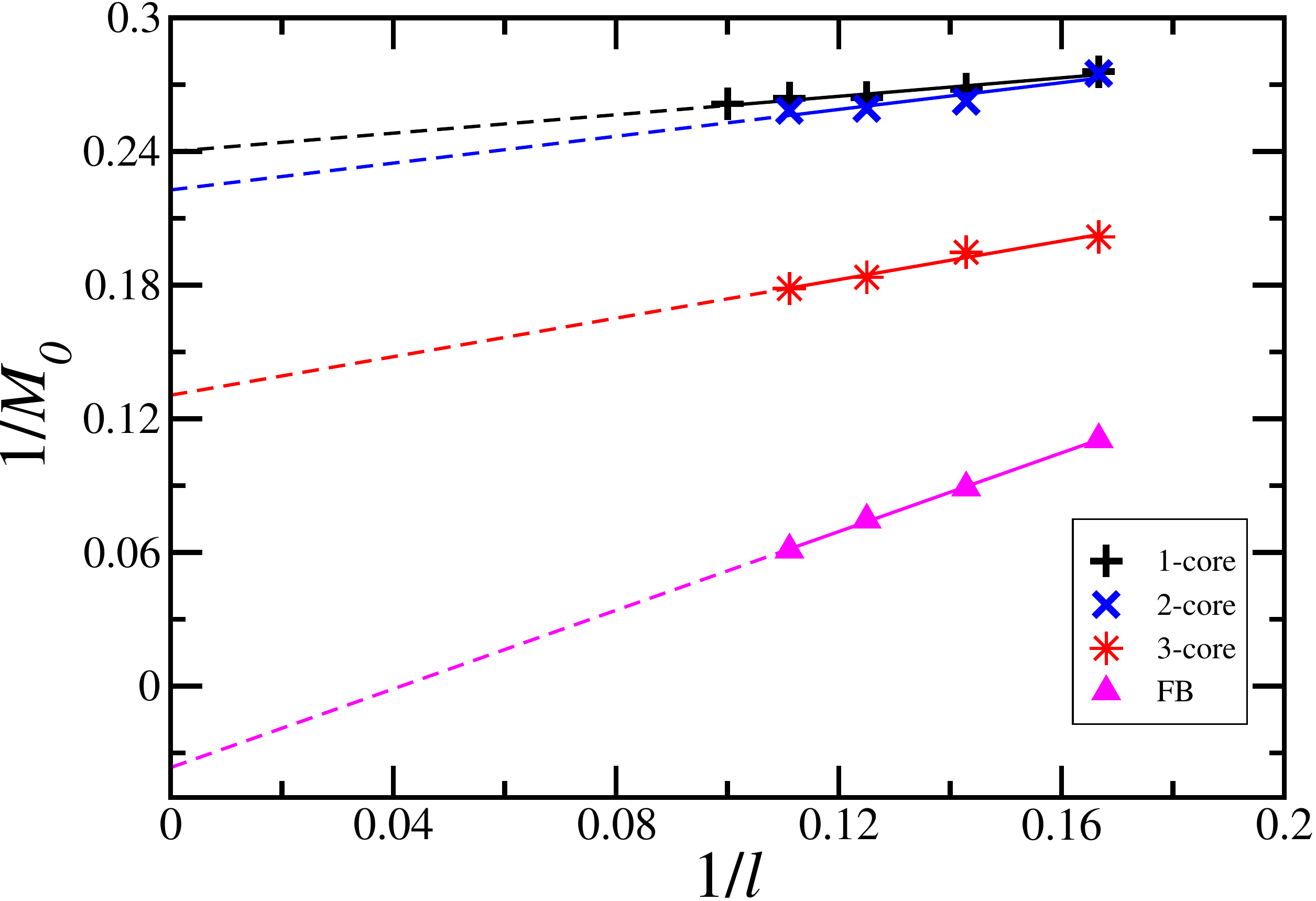}
	\end{center}
	\caption{Inverse slope of the crossing probability at the
          inflection point, $1/M_{0}$ as a function of $1/l$ for the
          different percolation models on the $\{3,7\}$
          lattice. For 1-core model $1/M_{0}$ tends to 0.240, for
          2-core to 223, for 3-core to 0.131, and for force-balance (FB) to -0.036 indicating $M_{0}$ is tending to $\infty$ as $l$ tends to $\infty$.}
	\label{CrossProb2}
\end{figure}

Since $k=1$-core percolation removes only isolated occupied sites, it
is essentially ordinary percolation. We should, therefore, observe
this finite slope intermediate region in the crossing probability as
the number of layers tends towards infinity. This finite slope region
has indeed been observed in Ref.~\cite{Ziff_Hyperbolic} for
$k=0$-core, or ordinary, percolation. Fig.~\ref{CrossProb1} presents
the crossing probability for all four percolation models. To check for
the existence of the intermediate region in $R(p)$ we extract its
maximum slope $M_{0}$ near the inflection point. We then plot the
inverse of this slope as a function of the $1/\ell$ and extrapolate to
the number of layers, $\ell$, going to infinity limit. The results are
illustrated in Fig.~\ref{CrossProb2}. The inverse of the slope,
$1/M_{0}$ tends to similar values for $k=1$-core and $k=2$-core
models. For $k=1$-core, it tends to 0.240 and 0.223 for the $k=2$-core
model. Meanwhile, $1/M_{0}$ tends to 0.131 for $k=3$-core model. 

The
fact that the inverse of the slope tends to -0.036 for the
force-balance model, which is much closer to zero than the three other
models, is an indication that the slope tends to infinity at the
transition. Then force-balance model would then exhibit just two
phases, one with no percolating cluster and the other with one
percolating cluster as ordinary percolation on Euclidean lattices. To
make a more rigorous case for the discontinuity of the crossing 
probability for the force-balance model, we analyze the tendency of
the inverse of the slope $1/M$ against $1/\ell$ for points located on
the intersection with the lines $R(p)=c, c \in \mathbb{R}$. For $c=0.3, 0.4, 0.5, 0.6,
0.7$, the inverse of the slope tends to a
negative value that is close to zero. This confirms the argument that $R(p)$ is discontinuous 
for the force-balance model, and, consequently, there should be just two phases for this model.

The suggestion of a finite slope regime of $R(p)$ for all three
$k$-core percolation models suggests that there is an intermediate
phase for all these models.  In other words, all three models behave
similarly to ordinary percolation. Of course, we have empirically
chosen a function to implement the extrapolation.  In
Ref.~\cite{Ziff_Hyperbolic}, the maximum slope $M$ as a function of
$N^{-0.7}$, where $N$ is the number of vertices in the tessellation,
was used.  We also tested different slightly extrapolation functions and our results remain unchanged in terms of the interpretation.

\subsection{Order parameter}

\begin{figure}[hb!]
	\centering
	\subfigure{	
		\includegraphics[width=0.3\textwidth]{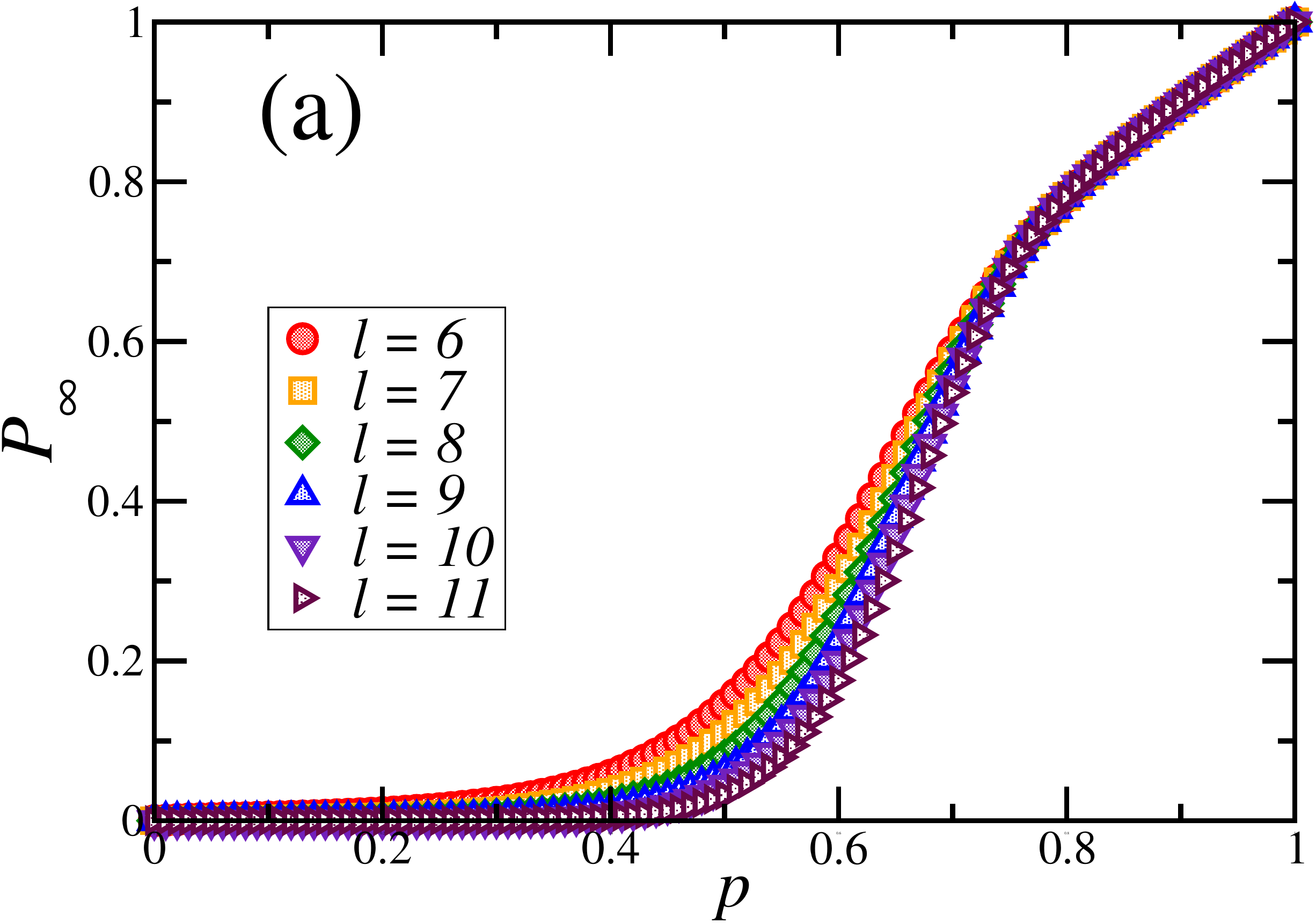}
		\label{OrderPar1:1}	}

	\subfigure{	
		\includegraphics[width=0.3\textwidth]{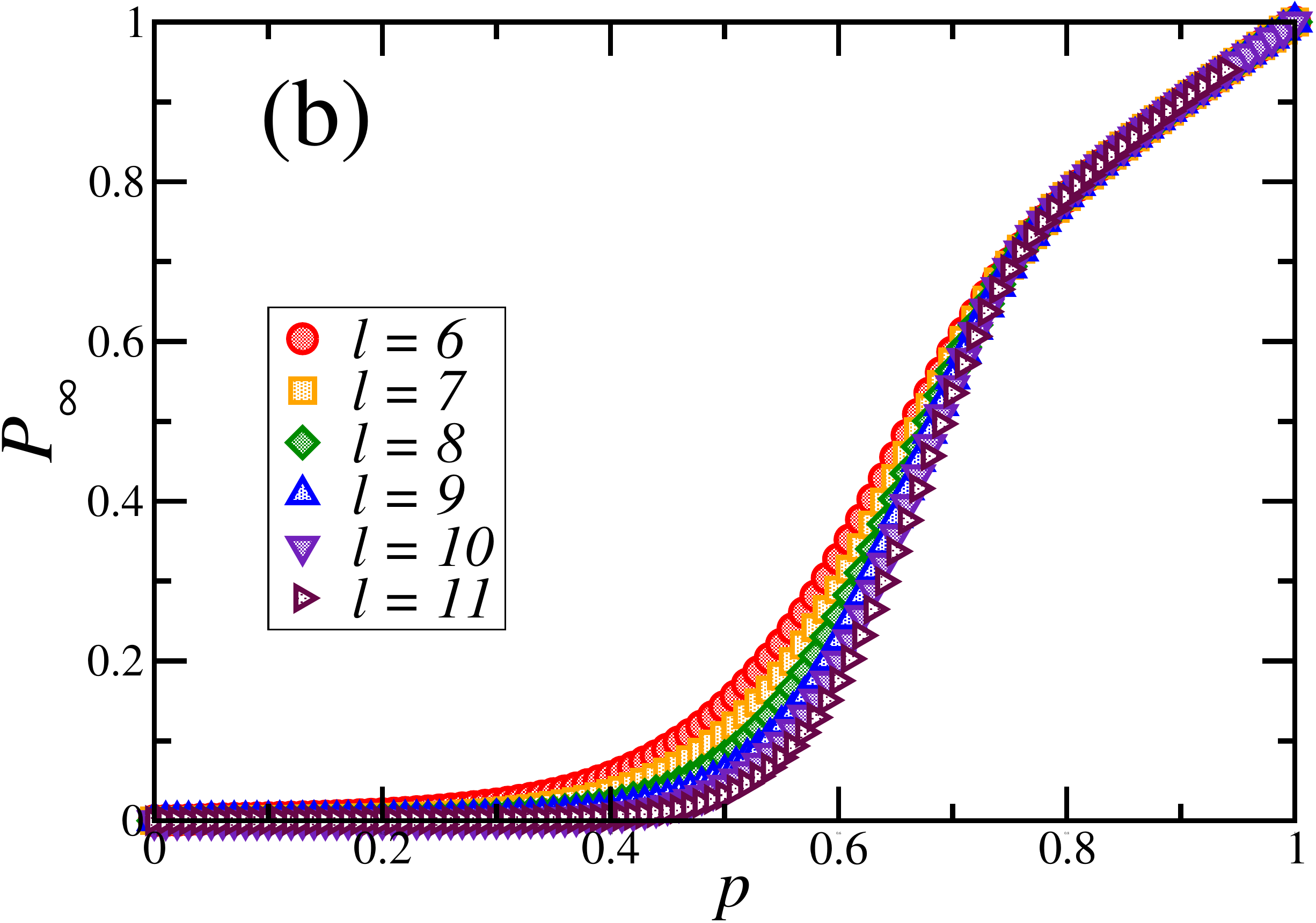}
		\label{OrderPar1:2}}	

	\subfigure{	
		\includegraphics[width=0.3\textwidth]{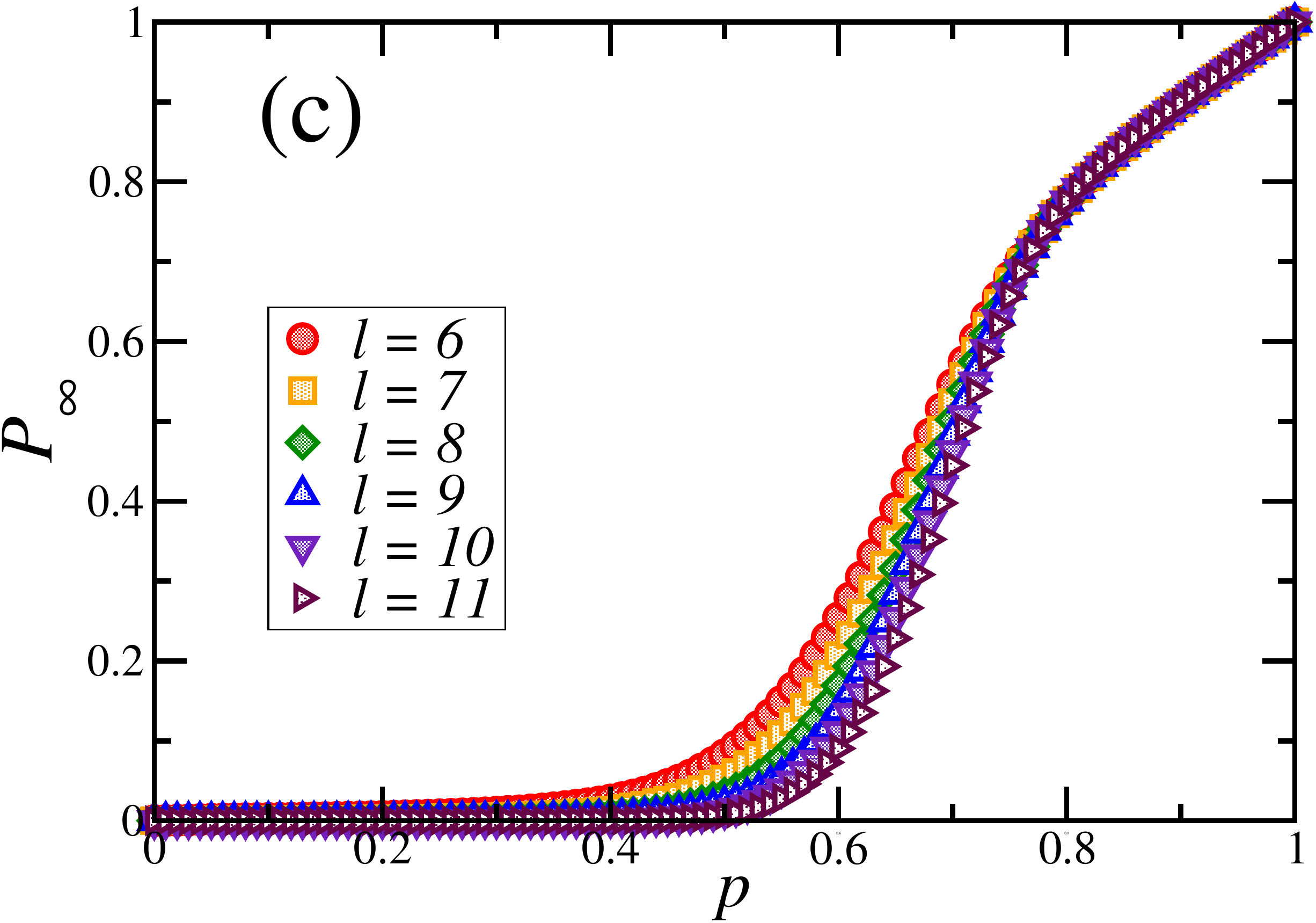}
		\label{OrderPar1:3}	}

	\subfigure{	
		\includegraphics[width=0.3\textwidth]{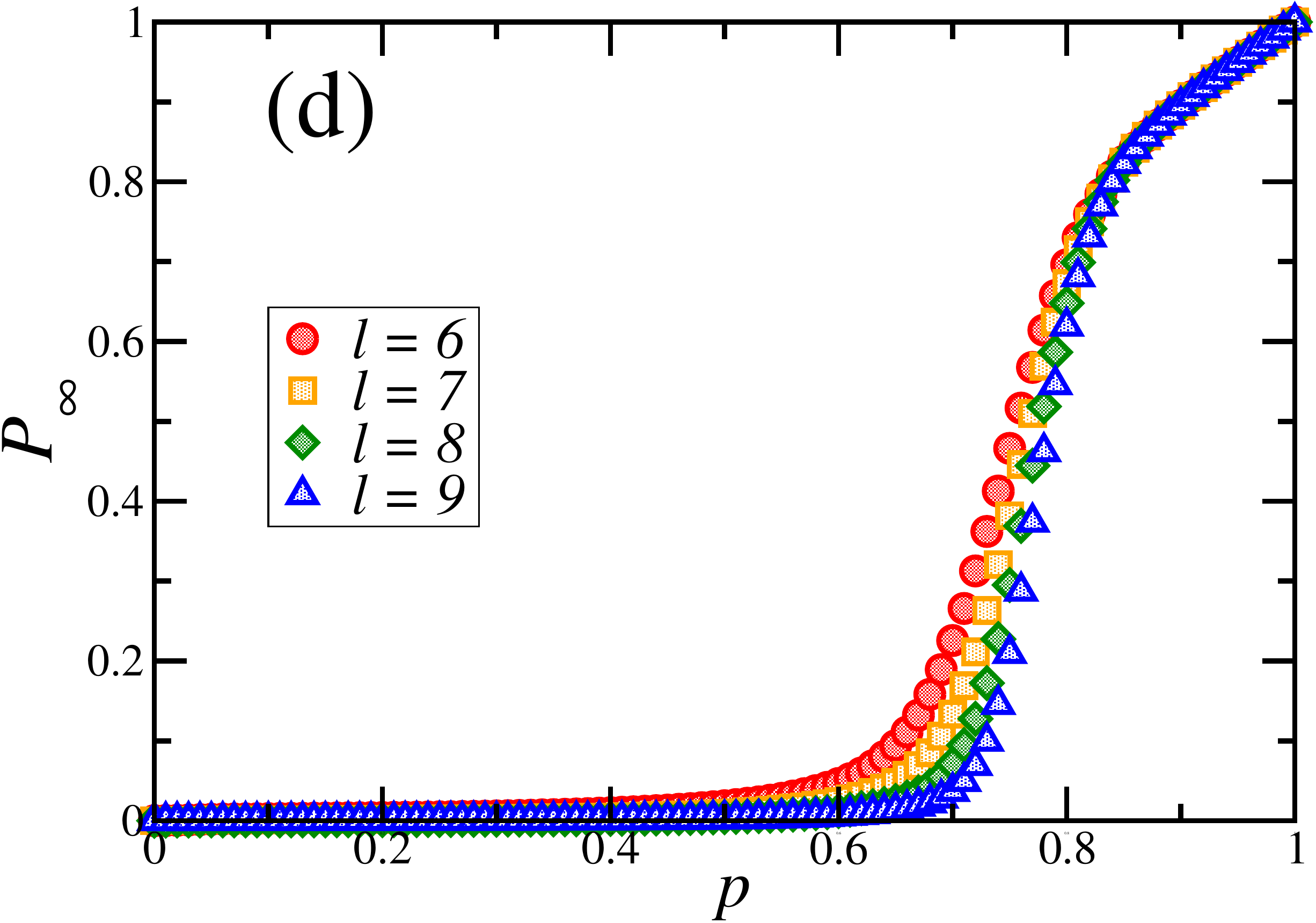}
		\label{OrderPar1:4}}	
	\caption{The fractional size of the largest cluster $P_{lc}$
          for the different percolation models on the $\{3,7\}$
          lattice: (a) $k=1$-core, (b) $k=2$-core, (c) $k=3$-core, 
          (d) force balance.}	
	\label{OrderPar1}
\end{figure}   

\begin{figure}[hb!]
	\begin{center}
\vspace{0.5cm}
		\includegraphics[width=0.4\textwidth]{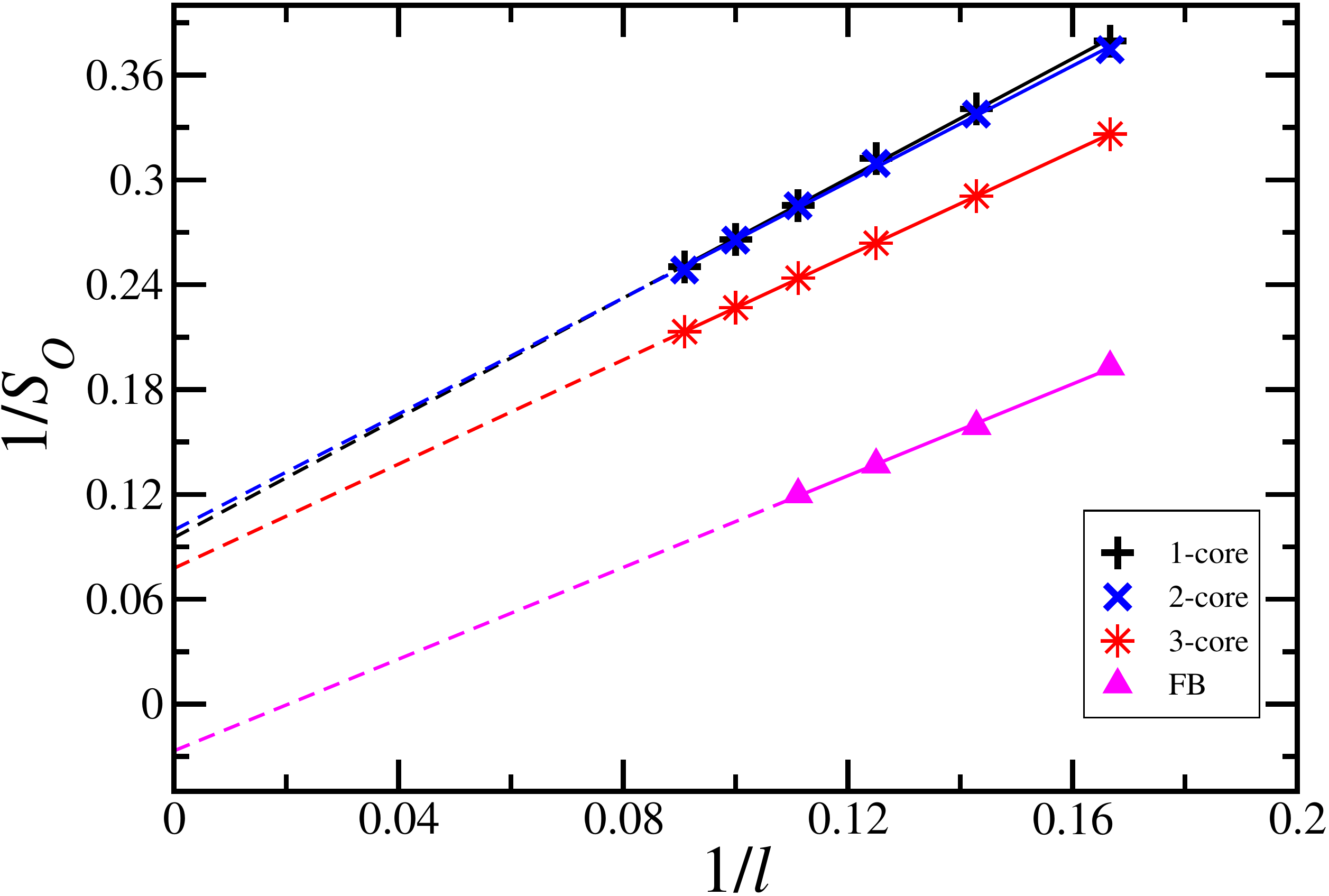}
	\end{center}
	\caption{The inverse of the maximum slope of $P_{\infty}$ as a
          function of $\ell$ for the different models percolation on the $\{3,7\}$ lattice.}
	\label{OrderPar2}
\end{figure}

\begin{figure}[hb!]
	\begin{center}
	\vspace{0.5cm}
		\includegraphics[width=0.4\textwidth]{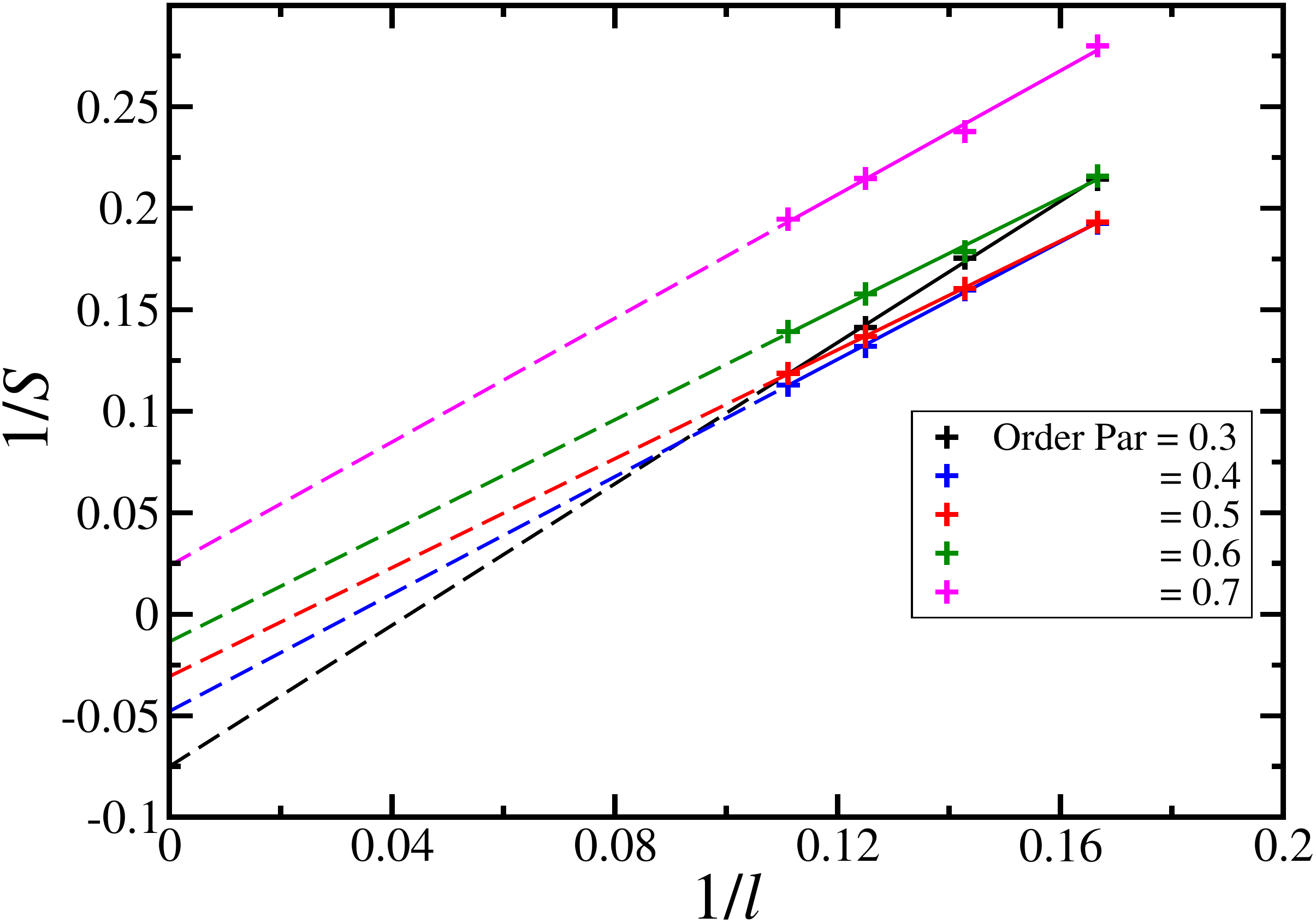}
			\end{center}
	\caption{Inverse slope of $P_{\infty}$, $1/S$, tendency on $1/l$ for points on the line $P_{\infty}=c$ and $c=0.3,0.4,0.5,0.6,0.7$, for the 
	force-balance model on the $\{3,7\}$ lattice.}
	\label{OrderPar_CB_dis}
\end{figure}

For ordinary percolation on Euclidean lattices the order parameter, $P_{\infty}$, is a continuous function of $p$ ~\cite{percolation}. Since $k=1,2$-core models are equivalent to unconstrained percolation in terms of the transition, they should behave similarly. While the order parameter in $k=3$-core on the Bethe lattice jumps discontinuously at the transition ~\cite{chalupa}, on Euclidean lattices it does not. For force-balance percolation on two- and three-dimensional Euclidean lattices, the order parameter jumps discontinuously at the transition ~\cite{schwarz2}. We present $P_{\infty}(p)$ for different layer numbers for the four different models on the $\{3,7\}$ tessellation in Fig.~\ref{OrderPar1}. 
Since any difference between the curves is not clear by eye, we perform a similar extrapolation to what was used for the study of $R(p)$. We measure the maximum slope of each curve and plot the inverse of the maximum slope, $1/S_{0}$ with respect to $1/l$. We found that the $1$-core and $2$-core models have a very similar limiting value, $1/S_0=0.0953$ and $1/S_0=0.0996$, respectively (see Fig.~\ref{OrderPar2}). $1/S_{0}$ tends to $0.0686$ for the $3$-core case, which is different than the previous two cases, but still non-zero. The $k$-core models may indeed be continuous phase transitions for the $\{3,7\}$ tessellation. For the force-balance model, the same extrapolation method yields tends to a negative value as shown in Fig.~\ref{OrderPar2} that is very close to zero. This result may indicate that force-balance percolation belongs to a discontinuous phase transition. This result is expected since it is discontinuous on Euclidean lattices as well. 
To make a more clear statement about the discontinuity of the force-balance transition, we analyze the behavior of the derivative for points on a line $P_{\infty}=c, c \in (0,1)$. We present the extrapolation of the inverse of this derivative $1/S$ versus the inverse number of layers $1/l$ for the values $c=0.3,0.4,0.5,0.6,0.7$ in Fig.~\ref{OrderPar_CB_dis}. We conclude that as $1/S$ is tending to negative values very close to zero for several values of the constant $c$, then $P_{\infty}$ is discontinuous implying that force-balance model is discontinuous on the tessellation $\{3,7\}$.

Note that it is interesting that $3$-core model is exhibiting a
continuous transition given that Sausset {\it et al.}~\cite{sausset2}
argue that the transition should be discontinuous. However, they do not study the tessellation $\{3,7\}$ and the criteria they used for a percolating cluster is one containing the central site and reaching the boundary, which is different from the criteria we use as we demand the percolating cluster to connect the two opposite boundary quarters sites.

\subsection{Culling time}
The culling time is the number of sweeps through the lattice to complete the culling/removal process for those occupied sites not 
obeying the respective constraints. On Euclidean lattices, the culling time for $k=3$-core and force-balance percolation increases 
near the percolation transition due to an increasing lengthscale in the distance over which the removal of one occupied site triggers the 
removal of other occupied sites. \vspace{0.4cm}

\begin{figure}[hb!]
	\centering
	\subfigure{	
		\includegraphics[width=0.3\textwidth]{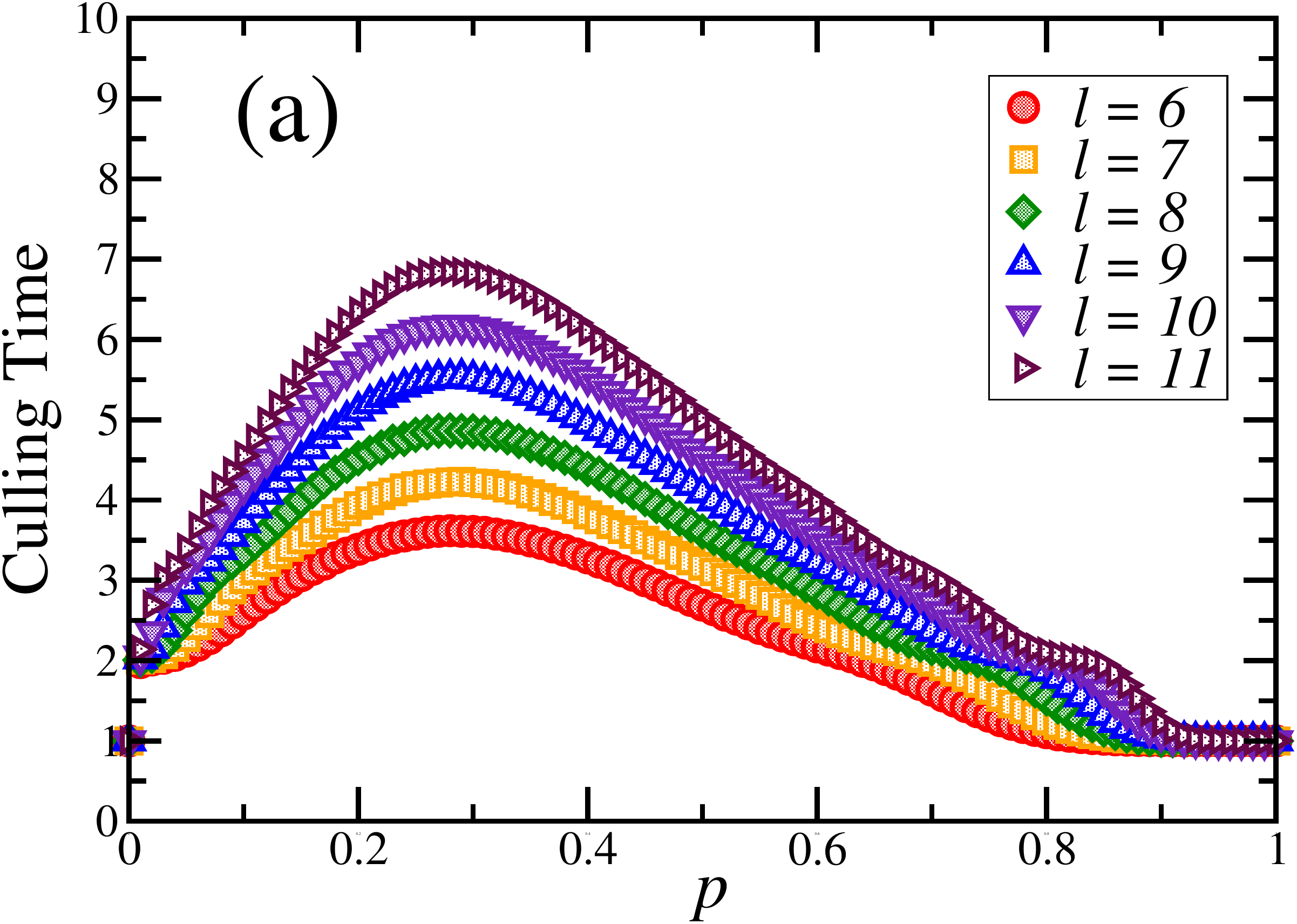}

		\label{fig9:2}	}

	\subfigure{	
		\includegraphics[width=0.3\textwidth]{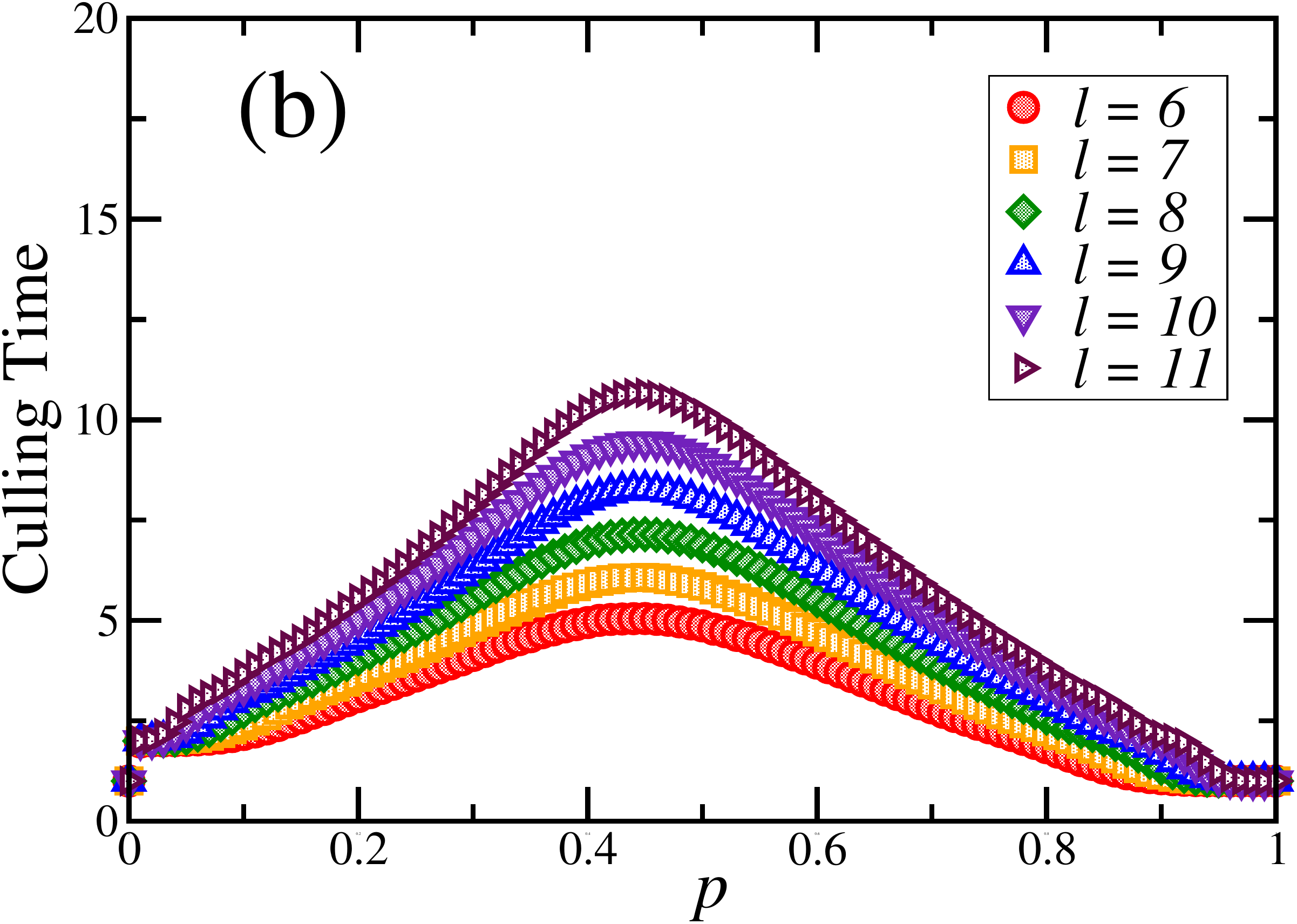}
		\label{fig9:3}
	}

	\subfigure{	
		\includegraphics[width=0.3\textwidth]{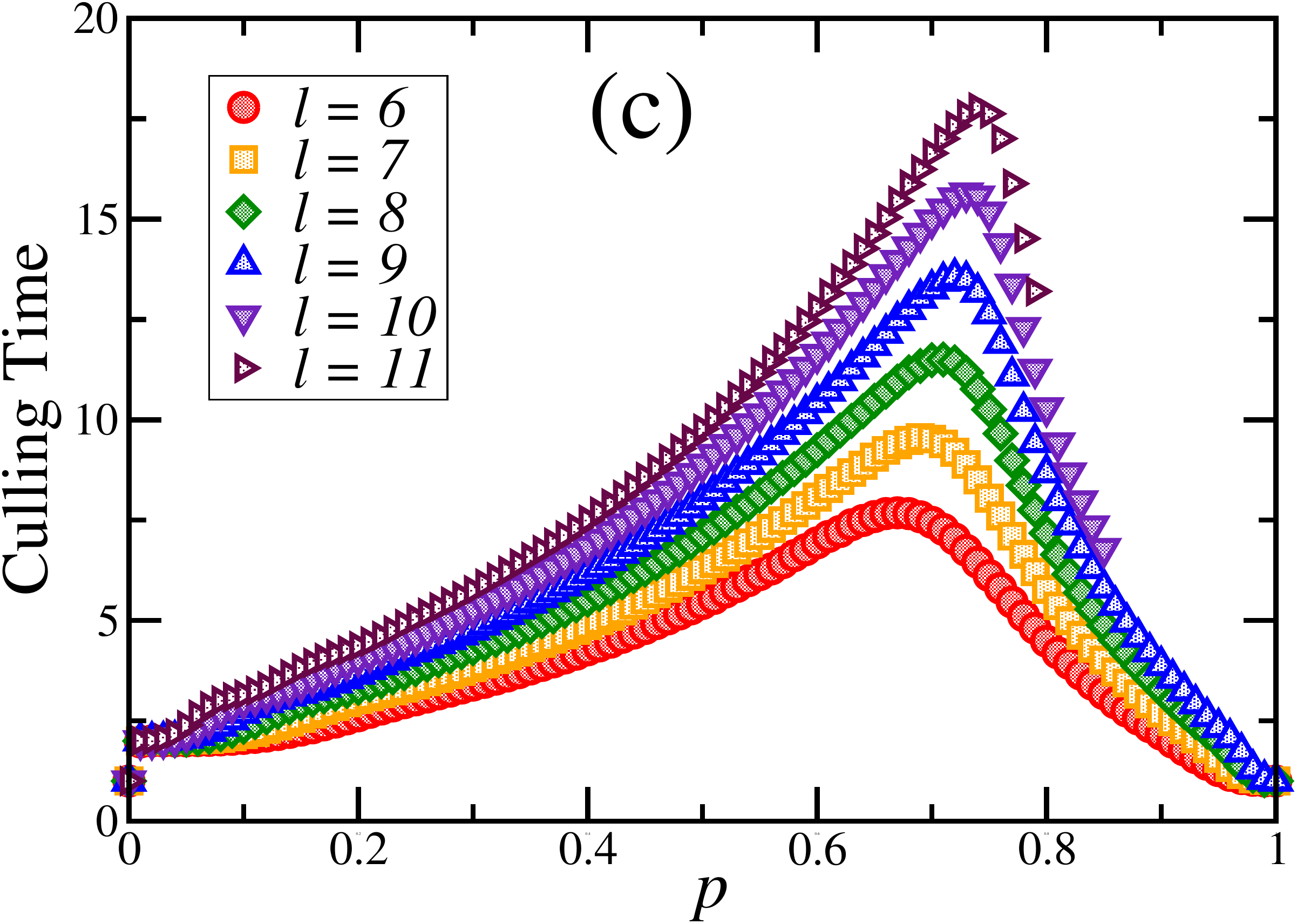}
		\label{fig9:4}	}
	\caption{Culling time for the different constraint percolation
          models for the $\{3,7\}$ lattice: (a) $k=2$-core, (b) $k=3$-core, (c)
          force balance. Each data set was averaged
          over 50,000 samples. }	
	\label{fig9}
\end{figure}  
	
\begin{figure}[h!]
	\begin{center}
	\vspace{0.6cm}
		\includegraphics[width=0.4\textwidth]{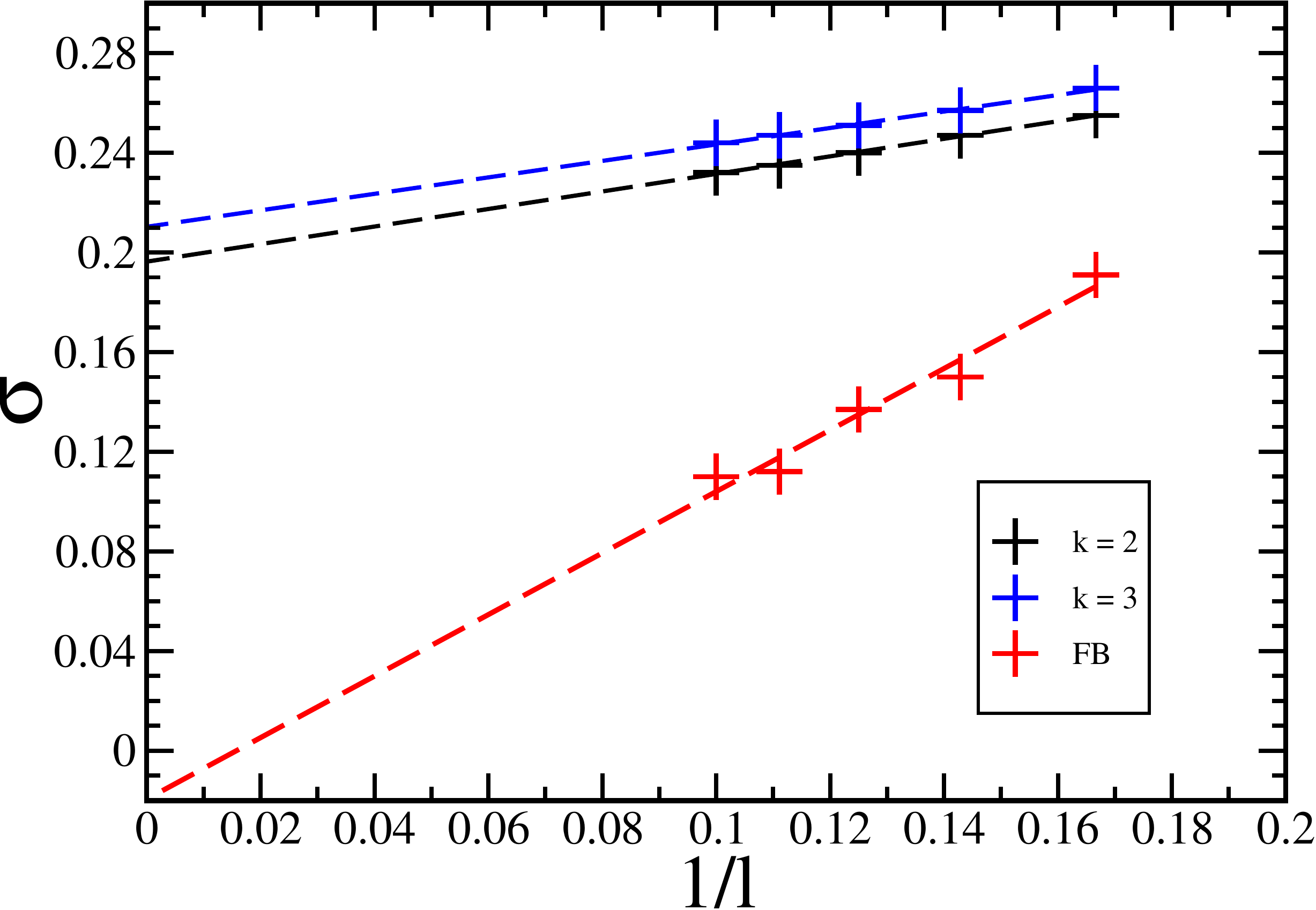}
			\end{center}
	\caption{Behavior of the width $\sigma$ vs $1/l$ the 2-core, 3-core and force balance models on the $\{3,7\}$ lattice.}
	\label{Width}
\end{figure}

In Fig.~\ref{fig9}, we observe the culling time for tessellation $\{3,7\}$, for $k=2,3$-core and 
force-balance models. Note that for $k=1$-core it just takes one sweep of the lattice to eliminate sites no satisfying the constraint so 
there is no diverging lengthscale.  According to Fig.~\ref{fig9}, there is a peak in the culling time $T$ as a function of $p$. Note that the 
position of the peak for the $k$-core models does not move and presumably yields an estimate for $p_l$. Accordingly, on $\{3,7\}$ for 
the $2$-core model, $p_{2l}\approx 0.28$, and for the $3$-core model $p_{3l}\approx 0.45$. However, for force-balance model the 
peak is increasing with the number of layers. We obtain the extrapolated $p_{FB}^*\approx 0.837$ when scaling $p_{FB}$ as $l^{-1}$. 
We approximate each curve to a gaussian function $f(x) = Ae^{-(x-x_{0})^2/\sigma^2}$ in a region close to the peak. The tendency of
$\sigma$ vs $1/l$ is illustrated in Fig.~\ref{Width}. Therefore, the
width $\sigma$ tends to a finite value for the $k-$core models, 0.196
for 2-core and 0.210 for 3-core, while it shrinks to zero for the force-balance model. Furthermore, the height of the peak tends to
infinity for all these peaks. A peak that remains broad in the
infinite system limit may be indicative of the two percolation
thresholds in the ordinary percolation model that appear to survive in the
$k=2$- and $k=3$-core models.

\subsection{Debate about $p_{u}$}

\begin{figure}[hb!]
	\centering
		\includegraphics[width=0.4\textwidth]{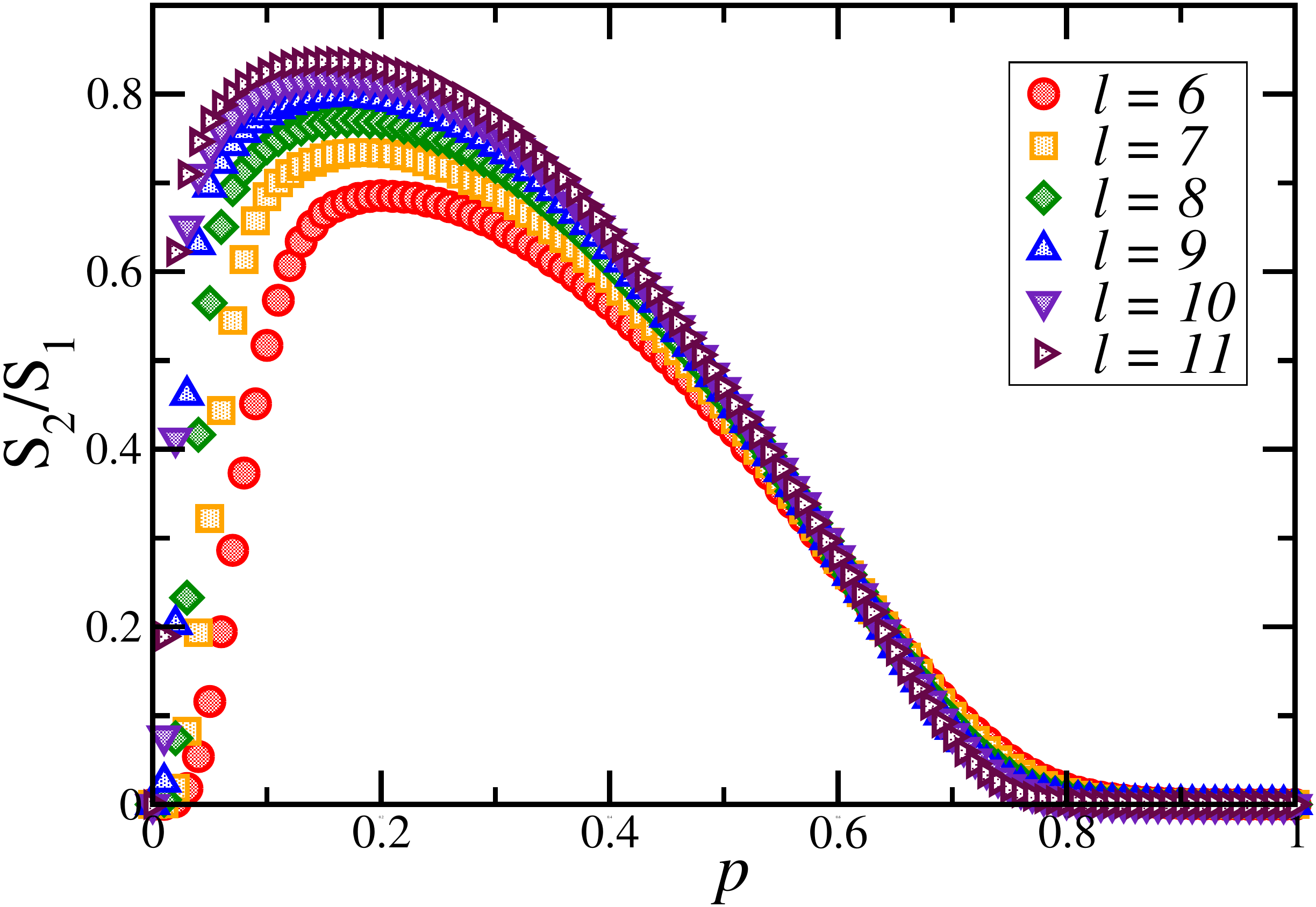}
	\caption{Ratio $S_2/S_1$ for $k=1$-core and for the tessellation $\{3,7\}$.}	
	\label{fig8}
\end{figure}

There exist three phases for ordinary percolation on a hyperbolic lattice~\cite{benjamini}. For $p<p_{l}$ there is no percolating 
cluster, for $p_{l}<p<p_{u}$ there are infinitely many percolating
clusters, and for $p_{u}<p$, the infinite number of percolating clusters 
join form one. There is no clear consensus, however, about how to numerically calculate $p_{l}$ and $p_{u}$ ~\cite{Baek3}. According 
to Ref.~\cite{Baek1}, $p_{l}$ can be measured as the probability above which there is a cluster connecting boundary points to the 
center. But $p_{u}$ can be measured in three different ways. The probability above which the ratio between the second biggest cluster 
and the biggest cluster, $S_{2}/S_{1}$, becomes negligible, or there is a finite fraction of the boundary points connected to the middle, 
or the probability at which the cluster size distribution $P(s)$ becomes power law. Furthermore, for calculating $p_{u}$, Ref.~\cite{Baek1} determines a way of 
finding $p_u$ by measuring the ratio $S_{2}/S_{1}$ between the second largest and largest clusters. The initial claim was that in the infinite 
limit such a curve will be discontinuous at some intersection point
(see their Fig. 4). However, in a more recent paper ~\cite{Baek3}, the
same authors state it could be the case that the curve is not discontinuous at this point, such as the curves for $R(p)$. In fact, according 
to Fig.~\ref{fig8} this seems to be the case here for the
$k=1$-core model (and for the other two $k$-core models as well). So we do not rely on this method any further. 
\vspace{0.5cm}

According to Ref.~\cite{Ziff_Hyperbolic}, $p_{l}$ and $p_{u}$ can be measured from the crossing probability $R(p)$, i.e. the probability 
of having a cluster going from one side of the lattice to the other. While this is the more straightforward measure, it would be good to find 
other measurements as a consistency check. It is important to note that there is a relationship between $p_{l}$ and $p_{u}$ on a lattice 
and its values on the dual lattice that are denoted as
$\overline{p_{l}}$ and $\overline{p_{u}}$, respectively. Such
relationship is given by
\begin{equation} \label{dualLattice}
p_l + \overline{p_{u}} = 1, \,\,\,\,\,\,\,\, \overline{p_{l}} + p_u = 1.
\end{equation}
And the dual lattice to $\{m,n\}$ is  $\{n,m\}$~\cite{Ziff_Hyperbolic}. As the measurement of $p_{l}$ is less controversial than the one 
for $p_{u}$ we can use Eq.~\eqref{dualLattice} to calculate $p_{u}$ by calculating $\overline{p_{l}}$ on the dual lattice. To estimate 
$p_l$ we search for the point at which the crossing probability is greater or equal than $10^{-4}$, similar to the procedure followed in 
Ref.~\cite{Ziff_Hyperbolic}. For these calculations, the data was averaged over 100000 runs and has large fluctuations.  We estimate 
$p_{l}$ for the $k$-core models on the tessellation $\{3,7\}$. For $k=1$-core, $p_{l}=0.20$; for $k=2$-core, $p_{l}=0.24$; for 
$k=3$-core, $p_{l}=0.37$. According to Eq. (\ref{dualLattice}), for ordinary percolation ($k=1$-core) on tessellation 
$\{7,3\}$, we should have $p_{u} = 0.80$. In order to estimate $p_{u}$ numerically (for $k=1 $-core model), we follow the procedure
outlined in Ref.~\cite{Ziff_Hyperbolic} stating that $p_{u}$ is the value of $p$ at which the ratio the crossing probability $R(p)$ 
becomes equal to one, for tessellations $\{3,7\}$ and $\{7,3\}$. Accordingly, the best estimate for $p_{u}$ for the tessellation $\{3,7\}$ 
is $p_{u}=0.73\pm0.02$ and for the tessellation $\{7,3\}$, $p_{u}=0.86\pm0.02$ which roughly satifies Eq. (\ref{dualLattice}).

\section{Discussion}

We have studied four constraint percolation models on mainly the
$\{3,7\}$ hyperbolic tessellation. Our data suggests that all three
$k$-core models exhibit similar behavior, thereby falling under the
universality class of ordinary percolation. This is not a surprise for
$k=2$-core percolation, which has been shown to behave similarly to 
ordinary percolation~\cite{Harris}. However, given the mixed
$k=3$-core percolation transition on Bethe lattices and, yet, the continuous phase
transition (should $p_c<1$) on Euclidean lattices for $k=3$-core, this
result is not obvious. In fact, earlier work ~\cite{sausset2} of $k=3$-core
percolation on hyperbolic lattices argued that the transition behaves
discontinuously (in terms of the order parameter).  In this earlier
work, it was also proven that $p_c<1$ for $k=3$-core percolation on
hyperbolic lattices of certain types, which is an important step in
that it constrains the analysis of the data.  However, there was no
proof of the discontinuity. Arguments and some numerical evidence were
presented, so our results do not contradict mathematics. 

Another interesting result is that the $k$-core models exhibit two
critical probabilities, $p_{l}$ and $p_{u}$, meanwhile the
force-balance model seems to exhibit just one critical
probability. This comes from the fact that the force-balance condition
constrains the spatial occupation of neighbors of an occupied site in
such a way that the cluster tends to expand in every direction. It
does not allow for the possibility of having several percolating clusters that do not overlap.

According to the behavior of the order parameter, $P_{\infty}$,
$k$-core models exhibit a continuous transition, while force-balance
is discontinuous, at least on the $\{3,7\}$
tessellation. Force-balance percolation is also discontinuous on
Euclidean lattices so the hyperbolic lattice does not change this property by the changing of the underlying geometry. 

The observation that the nature of the transition in $k=3$-core
percolation does not change from Euclidean lattices to hyperbolic
lattices may indicate that (the absence of) loops are important in driving the
transition towards a mixed one since on the Bethe
lattice there are no loops. In other words, $k=3$-core
percolation may be very sensitive to loops. A $1/d$ expansion for $k=3$-core percolation demonstrated that
the mixed nature of the transiton remained to order
$1/d^3$~\cite{HarrisSchwarz}. Of course, the loops are controlled
perturbatively in this $1/d$ expansion, which is not the case for the
hyperbolic tessellation. One must think about the effects of loops on
$k=3$-core percolation to better understand the nature of its transition in all geometries. 

JMS acknowledges support from NSF-DMR-1507938 and the Soft Matter
Program at Syracuse University.

\end{document}